\documentclass[traditabstract,longauth]{aa}
\usepackage{amssymb}
\usepackage{mathptmx}
\usepackage{natbib}
\usepackage[]{graphicx}
\usepackage{longtable}
\usepackage{lscape}

\begin{document}

\title{Kinematic signatures of AGN feedback in moderately powerful radio galaxies at z$\sim$2 observed with SINFONI\thanks{Based on observations carried out with the Very Large Telescope of ESO under Program IDs 084.A$-$0324 and 
085.A$-$0897, and at ATCA under Program ID C2604.}}
\author{C.~Collet\inst{1}, N.~P.~H.~Nesvadba\inst{1,2}, C.~De~Breuck\inst{3}, M.~D.~Lehnert\inst{4}, P.~Best\inst{5}, J.~J.~Bryant\inst{6,7,8}, R.~Hunstead\inst{6}, D.~Dicken\inst{1,9} and H.~Johnston\inst{6}}
\institute{Institut d'Astrophysique Spatiale, CNRS, Centre Universitaire
  d'Orsay, Bat. 120$-$121, 91405 Orsay, France 
\and 
email: nicole.nesvadba@ias.u$-$psud.fr
\and
European Southern Observatory, Karl-Schwarzschild Strasse, Garching bei M\"unchen, Germany
\and
Institut d'Astrophysique de Paris, CNRS \& Universit\'e Pierre et
Marie Curie, 98bis, bd Arago, 75014 Paris, France
\and
SUPA, Institute for Astronomy, Royal Observatory of Edinburgh, Blackford Hill, Edinburgh EH9 3HJ, UK
\and
Sydney Institute for Astronomy (SIfA), School of Physics, The University of Sydney, NSW 2006, Australia 
\and
Australian Astronomical Observatory, PO Box 915, North Ryde, NSW1670, Australia 
\and
ARC Centre of Excellence for All-sky Astrophysics (CAASTRO), Australia
\and
Laboratoire AIM Paris-Saclay, CEA/DSM/Irfu, Orme des Merisiers, Bat 709, 91191 Gif sur Yvette, France
}
\titlerunning{Moderately powerful high-z radio galaxies}
\authorrunning{Collet et al.}  \date{Received / Accepted }

\date{Accepted . Received ; in original form }

\abstract{
Most successful galaxy formation scenarios now postulate that the
intense star formation in massive, high-redshift galaxies during their
major growth period was truncated when powerful AGNs launched
galaxy-wide outflows of gas that removed large parts of the interstellar medium.
SINFONI imaging spectroscopy of the most powerful radio galaxies at
z$\sim$2 show clear signatures of such winds, but are too rare to be
good representatives of a generic phase in the evolution of all
massive galaxies at high redshift. Here we present SINFONI
imaging spectroscopy of the rest-frame optical emission-line gas in
12 radio galaxies at redshifts $\sim 2$. Our sample spans a range in
radio power that is intermediate between the most powerful radio
galaxies with known wind signatures at these redshifts and vigorous
starburst galaxies, and are about two orders of magnitude more
common than the most powerful radio galaxies. Thus, if AGN feedback
is a generic phase of massive galaxy evolution for reasonable values
of the AGN duty cycle, these are just the sources where AGN feedback
should be most important.

Our sources show a diverse set of gas kinematics ranging from regular
velocity gradients with amplitudes of $\Delta v=$200$-$400 km
s$^{-1}$ consistent with rotating disks to very irregular kinematics
with multiple velocity jumps of a few 100 km s$^{-1}$.  Line widths are
generally high, typically around FWHM$=$800 km s$^{-1}$, more similar to 
the more powerful high-z radio galaxies than mass-selected samples of
massive high-z galaxies without bright AGNs, and consistent with the
velocity range expected from recent hydrodynamic models. A broad
H$\alpha$ line in one target implies a black hole mass of a few $10^9$
M$_{\odot}$. Velocity offsets of putative satellite galaxies near a
few targets suggest dynamical masses of a few $10^{11}$ M$_{\odot}$ for
our sources, akin to the most powerful high-z radio galaxies. Ionized
gas masses are 1$-$2 orders of magnitude lower than in the most
powerful radio galaxies, and the extinction in the gas is relatively
low, up to A$_V\sim$2~mag. The ratio of line widths, $\sigma$, to bulk
velocity, $v$, is so large that even the gas in galaxies with regular
velocity fields is unlikely to be gravitationally bound. It
is  unclear, however, whether the large line widths are due to turbulence or
unresolved, local outflows as are sometimes observed at low redshifts.  We
compare our sources with sets of radio galaxies at low and high
redshift, finding that they may have more in common with gas-rich
nearby radio galaxies with similar jet power than with the most
powerful high-z radio galaxies. Comparison of the kinetic energy with
the energy supply from the AGNs through jet and radiation pressure
suggests that the radio source still plays a dominant role for
feedback, consistent with low-redshift radio-loud quasars.}


\maketitle

\section{Introduction}
\label{sec:introduction}

It is now widely accepted that the supermassive black holes residing
in the vast majority of early-type galaxies and bulges
\citep[e.g.,][]{yu02} can have a sizeable impact on the evolution of
their host galaxies. Semi-analytic models require a strong source of
energy to balance the overcooling of gas onto dark matter halos and to avoid 
a strong excess in baryonic mass and star formation in galaxies
compared to observations. The discrepancy is strongest at the
high-mass end of the galaxy mass function, which is dominated by
early-type galaxies \citep[e.g.,][]{Benson2003}, whose mass and
structural properties appear closely related to the mass of their
central supermassive black hole \citep[e.g.,][]{tremaine02}.

Powerful AGNs release approximately the equivalent of the binding energy of a
massive galaxy during their short activity period, either in the form
of radiation, or through jets of relativistic particles, or both. They
are thus in principle able to offset the excess cooling out to the
highest galaxy masses \citep[][]{silk98}. If sufficiently large parts
of this energy are deposited in the interstellar medium of the host
galaxy, they may drive winds \citep[][]{dimatteo05} or turbulence
\citep[][]{Nesvadba2011c}. The mechanisms that cause this,
 however, are still not very well understood. Even very basic questions,
e.g., whether feedback is dominated by radio jets or the bolometric
energy output of radiatively efficient accretion disks, are still
heavily debated in the literature. There is clear observational
evidence  that jets can perturb the ISM strongly even at kpc
distance from the nucleus, while the observational evidence for winds driven
by quasar radiation over kpc scales is still mixed
\citep[e.g.,][]{husemann13,liu13,CanoDiaz2012,Harrison2012}. Hydrodynamic
models of radio jets are now finding deposition rates of kinetic
energy from the jet into the gas that are broadly consistent with
observations \citep[][]{Wagner2012}.

Following the most popular galaxy evolution models, AGN feedback
should have been particularly important in the early evolution of
massive galaxies at high redshift, where AGN-driven winds may have
blown out the remaining gaseous reservoirs that fueled the main phase
of galaxy growth, inhibiting extended periods of subsequent star
formation from the residual gas.  Whereas the jets of powerful radio
galaxies in the local Universe are known to affect the gas locally
within extended gas disks \citep[``jet-cloud interactions'',
  e.g.,][]{tadhunter98, vanbreugel85}, it was found only recently  that
outflows driven by the most powerful radio jets in the early universe
at z$\sim$2 can encompass very high gas masses, up to about $10^{10}$
M$_{\odot}$ in the most powerful radio galaxies at high redshift
\citep[][]{Nesvadba2006,Nesvadba2008}. This is similar to the typical
total gas masses in massive, intensely star-forming, high-redshift
galaxies of a few $10^{10}$ M$_{\odot}$ \citep[e.g.,][]{greve05,
  tacconi08}.  This gas is strongly kinematically perturbed with FWHM
up to $\sim2000$ km s$^{-1}$ and high, abrupt velocity gradients of
similar amplitude, consistent with the expected signatures of vigorous
jet-driven winds.

Galaxies with extreme radio power at 1.4~GHz of up to
P$_{1.4}=$few$\times 10^{29}$ W Hz$^{-1}$ like those studied by
\citet{Nesvadba2006,Nesvadba2008} are, however, very rare, which raises
the question of the impact that AGNs may have on their host galaxies
when their radio power is significantly lower. The present paper gives
a first answer to this question. It is part of a systematic study of
the warm ionized gas in 49 high-redshift radio galaxies at z$\sim$2
with SINFONI, which span three decades in radio power and two decades in
radio size. Our sources cover the lower half of the radio power of
this sample, P$_{1.4}=$few$\times 10^{26-27}$ W Hz$^{-1}$ at 1.4~GHz
in the rest frame. Toward lower radio power, contamination from the
non-thermal radio continuum of vigorous starbursts becomes
increasingly important. The high-redshift radio luminosity function of
\citet{Willott2001} and \citet{Gendre2010} suggests that such galaxies
are factors of 100 more common than the very powerful radio sources,
with co-moving number densities on the  order of a few $10^{-7}$ Mpc$^{-3}$,
sufficient to represent a short, generic phase in the evolution of
massive galaxies, as we will argue below in \S\ref{sec:ensemble}.

The organization of the paper is as follows. In \S\ref{sec:sample} we
present our sample and in
\S\ref{sec:observationsAndDataReduction_NVSS} our SINFONI NIR imaging
spectroscopy and ATCA centimeter continuum observations, and the
methods with which we reduced these data. In
\S\ref{s:PresentationOfOurAnalysisTools} we present our 
methods of analysis and the results for each individual target, before presenting the
overall results drawn from our sample in
\S\ref{sec:ensembleproperties}. In \S\ref{ss:agnproperties} we discuss
the AGN properties, and in
\S\ref{s:additionalLineEmittersTracersOfEnvironmentDynamicalProbes} we
use additional line emitters near our radio galaxies to estimate
dynamical masses of our high-redshift radio galaxies (HzRGs). In \S\ref{sec:results.globalproperties}
we compare our data with other classes of HzRGs before discussing the
implications of our results for AGN feedback. We argue in
\S\ref{sec:ensemble} that sources similar to those studied here may well
be a representative subset of massive high-redshift galaxies overall,
seen in a short but important phase of their evolution, and we summarize
our results in \S\ref{ssec:summary}.

Throughout our analysis we adopt a flat cosmology with H$_0$ = 70 km
s$^{-1}$ Mpc$^{-1}$, $\Omega_{\Lambda}$=0.7, and $\Omega_{M}=0.3$.

\section{Sample}
\label{sec:sample}

Our sources have a radio power of a few 10$^{27-28}$ W Hz$^{-1}$ at
500~MHz in the rest frame, about 2 to 3 orders of magnitude fainter
than the most powerful high-redshift radio galaxies known, which reach
up to nearly $1\times 10^{30}$ W Hz$^{-1}$ \citep[][ and references
  therein]{miley08}, but powerful enough to safely neglect
contamination from intense star formation. For comparison, an
intensely star-forming HyLIRG (hyper-luminous infrared galaxy) with
far-infrared luminosity $L_{FIR}=1\times 10^{13} L_{\odot}$
 would produce a rest-frame 1.4~GHz radio power of
$10^{25.0}$ W Hz$^{-1}$, assuming a far-infrared-to-radio luminosity
ratio of 2.0, as found for high-redshift submillimeter galaxies
\citep[e.g.,][]{vlahakis07, seymour09, thomson14}, and a radio
spectral index typical of star formation of $\alpha=-0.7$ to
$-0.8$. This is very similar to the steep spectral indices $\alpha
\sim -1.0$ that are characteristic of high-redshift radio galaxies,
making it even more difficult to disentangle the contribution of AGNs
and star formation to lower-power radio sources than those studied
here. In spite of their faintness relative to other high-redshift
radio galaxies, the radio power of fainter sources in this present
study is nonetheless comparable to that of the most powerful radio
galaxies known at low redshift \citep[e.g.,][]{tadhunter93}.

Our targets come from two different surveys. One is the southern
sample of 234 distant radio galaxies of \citet{Broderick2007},
\citet[][]{Bryant2009a}, \citet[][]{Bryant2009b}, and Johnston et
al. (in prep.), which we refer to hereafter as the ``MRCR-SUMSS''
sample. The other is the sample of radio galaxies within the fields of
the ESO imaging survey (EIS) by \citet{Best2003CENSORS} and
\citet{Brookes2006CENSORS, Brookes2008CENSORS}, which we call the
``CENSORS'' sample (``Combined EIS-NVSS Survey Of Radio Sources'').

The MRCR-SUMSS sources have steep radio spectral indices
$\alpha_{408-843} \le -1.0$ between 408~MHz and 843~MHz, and fluxes at
408~MHz $S_{408} \ge 200$~mJy. From this catalogue, we selected 12
moderately low-power sources at z~$\ge$~2 with $P_{1.4}$~=~few$\times
10^{27}$~W~Hz$^{-1}$. They have radio sizes between
$\sim$~2\arcsec\ and 24\arcsec\ at 1.4~GHz frequency, a typical range
of radio sizes of powerful HzRG.

The six galaxies with the lowest radio power
($\mathcal{P}_{1.4} \sim 10^{26}$~W~Hz$^{-1}$) come from
the CENSORS survey. This catalogue of 150 radio galaxies results from
cross-matching the ESO Imaging Survey (EIS) patch D with the NVSS
radio survey \citep[][]{best03}.  Radio sources detected in the NVSS were
re-observed at 1.4~GHz with the VLA in the BnA configuration at a
spatial resolution of 3\arcsec$-$4\arcsec, compared to the initial
spatial resolution of the NVSS of 45\arcsec, which is complete down to
7.2~mJy. This made it possible to study the structure of the radio
sources and to identify the most likely rest-frame optical counterparts
of 102 sources. Optical spectroscopy provided redshifts of 81 sources
\citep{Brookes2008CENSORS}.  Among these, we selected six sources
with a radio power of a few $\times 10^{26}$ W Hz$^{-1}$ and appropriate
redshifts for ground-based NIR follow-up spectroscopy. Three have extended
radio morphologies and three have compact, unresolved radio cores.

\section{Observations and data reduction}
\label{sec:observationsAndDataReduction_NVSS}

\subsection{Near-infrared imaging spectroscopy}
\label{ss:SINFONIobservations_NVSS}

We observed all MRCR-SUMSS galaxies with the near-infrared imaging
spectrograph SINFONI \citep[][]{SINFONI2003} between late 2009 and
early 2010 under program ID 084.A-0324 at the ESO Very Large
Telescope. SINFONI observations of the CENSORS sources were carried
out between late 2010 and early 2011 under program ID 086.B-0571. All
data were taken in Service Mode under variable conditions.

SINFONI is an image slicer that operates between 1.1$\mu$m and
2.4$\mu$m. We used the seeing-limited mode with the largest available
field of view of 8\arcsec$\times$8\arcsec\ and a pixel scale of
250~mas. All data were taken with the $H+K$ grating which covers
wavelengths between 1.45~$\mu$m and 2.4~$\mu$m at a spectral resolving
power R~$\sim$~1500 ($\sim$~200~km s$^{-1}$). We observed each MRCR-SUMSS
galaxy for 180$-$230 min of on-source observing time (300-440~min for
the CENSORS sources), split into individual observations of
5~min. Most of our galaxies are smaller than the field of view. We
therefore adopted a dither pattern where the object is shifted between
two opposite corners of the field of view, which allows us to use two
subsequent frames for the sky subtraction and makes taking dedicated
sky frames unnecessary. The spatial resolution of our data is limited
by the size of the seeing disk, which is typically around
1.0\arcsec\ for both samples. The FWHM sizes of the PSFs of individual
targets are given in Table~\ref{tab:obslog}.
They are measured from a standard star
observed at the end of each hour of data taking.

Data reduction relies on a combination of IRAF \citep[][]{tody93} and
our own custom IDL routines \citep[e.g.,][]{nesvadba11a}. All frames
are dark-subtracted and flat-fielded. We then remove the curvature
from the spectra in each slit and put them onto a common wavelength
scale by using the bright night-sky lines superimposed on each frame,
 using only arc lamp spectra  to set the absolute wavelength scale. We
then sky subtract our data and rearrange them into three-dimensional
data cubes, which are then combined. To account for the variability of
the night sky we scale the total flux in each sky frame to the total
flux in each object frame, after masking the target. We use the
standard star observations to correct for telluric and instrumental
effects and to set the absolute flux scale.

In this analysis, we discuss the optical emission-line properties of 8
of the 12 MRCR-SUMSS galaxies we observed. Two of the other four,
NVSS~J004136$-$345046 and NVSS~J103615$-$321659, have continuum emission but no
line emission at the redshifts previously measured in the rest-frame
UV. Their redshifts are relatively high, z$=$2.6, placing [OIII] and
H$\alpha$ at wavelengths outside the near-infrared atmospheric
  bands, where the atmospheric transmission is below 10\%, and is
  strongly variable both in time and in wavelength. At the expected
  wavelength of H$\alpha$, the telluric thermal background is already
  a factor of $\sim 10$ greater than at 2.2$\mu$m. A third source,
NVSS~J233034$-$330009, was found to coincide with a foreground star
after our data had already been taken \citep[][]{Bryant2009b}. The
fourth source, NVSS~J210626$-$314003 shows a strong misalignment
between the radio source and the extended gas, and no gas along the
radio jet axis, which is very different from the other galaxies
presented here. This source has already been discussed by \citet{collet14}, so
 we do not describe its  characteristics in detail again, but we do
include it in the overall discussion of the properties of our sources.
For the CENSORS sources we focus our discussion on the three sources where
we detected line emission at the redshifts previously measured in the
rest-frame UV.

\subsection{Radio continuum observations}
\label{ss:radioObservations_NVSS}

We observed our MRCR-SUMSS and CENSORS sources in two runs on 2012 January
28 and February 02 with the Australia Telescope Compact Array (ATCA,
project C2604).  Observations were carried out simultaneously at 5.5
and 9.0 GHz using the Compact Array Broadband Backend, with bandwidths
of 2 GHz and channel widths of 1~MHz. The array configuration was 6A,
with baselines between 337 and 5939~m. For flux density and bandpass
calibration we observed PKS B1934$-$638 at the beginning and end of
each session. Poor phase stability was  due to heavy rain and high
humidity, and this significantly affected the signal-to-noise ratio.

Individual sources were observed in 13--15 five-minute snapshots spread
over 8.5 hours to ensure good {\it uv\,} coverage; an exception was
NVSS J144932$-$385657, which set early with only five snapshots
spanning 3 hours.  We did not obtain any new data for CENSORS 072
because we used incorrect coordinates from Brooks et
al.\ (2008). Standard data reduction was carried out in {\sc miriad}
(Sault et al. 1995). The ATCA observing log is given in Tab.~2 and
lists for each source the date of observation, total on-source
integration time, the secondary phase calibrator used, and the
synthesized beam at each frequency.

The data reduction was done with \textsc{Miriad}
\citep[][]{MIRIAD1995} in the standard way. We find typical beam sizes
of 4\arcsec$\times$1.5\arcsec\ at 5.5~GHz and
2.5\arcsec$\times$0.9\arcsec\ with position angles between $-$6$^\circ$
and 13$^\circ$ (except for NVSS~J144932$-$385657, where
PA~=~$-$40$^\circ$). Details are given in Tab.~\ref{tab:atcalog}.

\begin{figure}
\begin{center}
\includegraphics[width=0.8\textwidth]{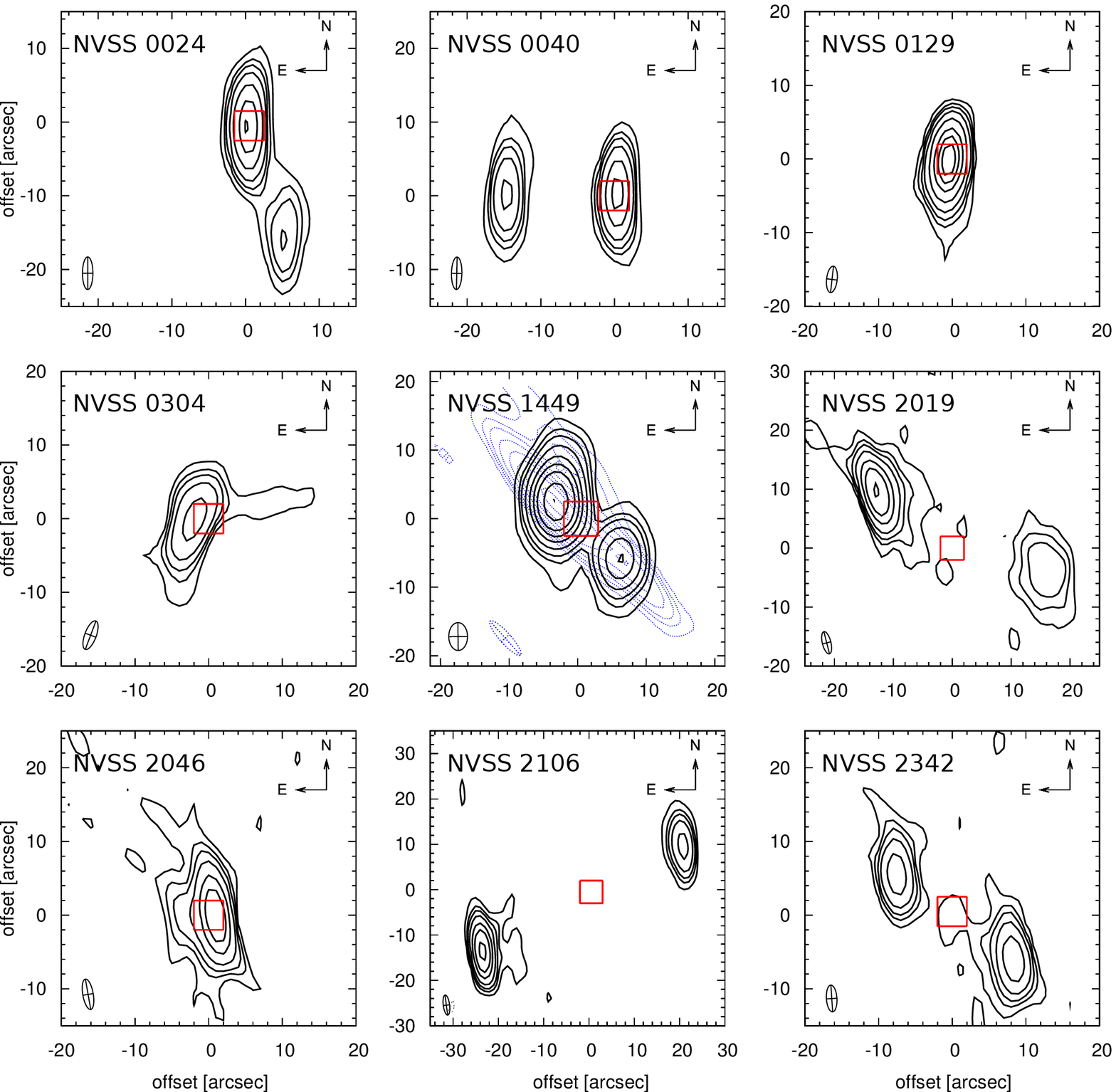}\\
\includegraphics[width=0.8\textwidth]{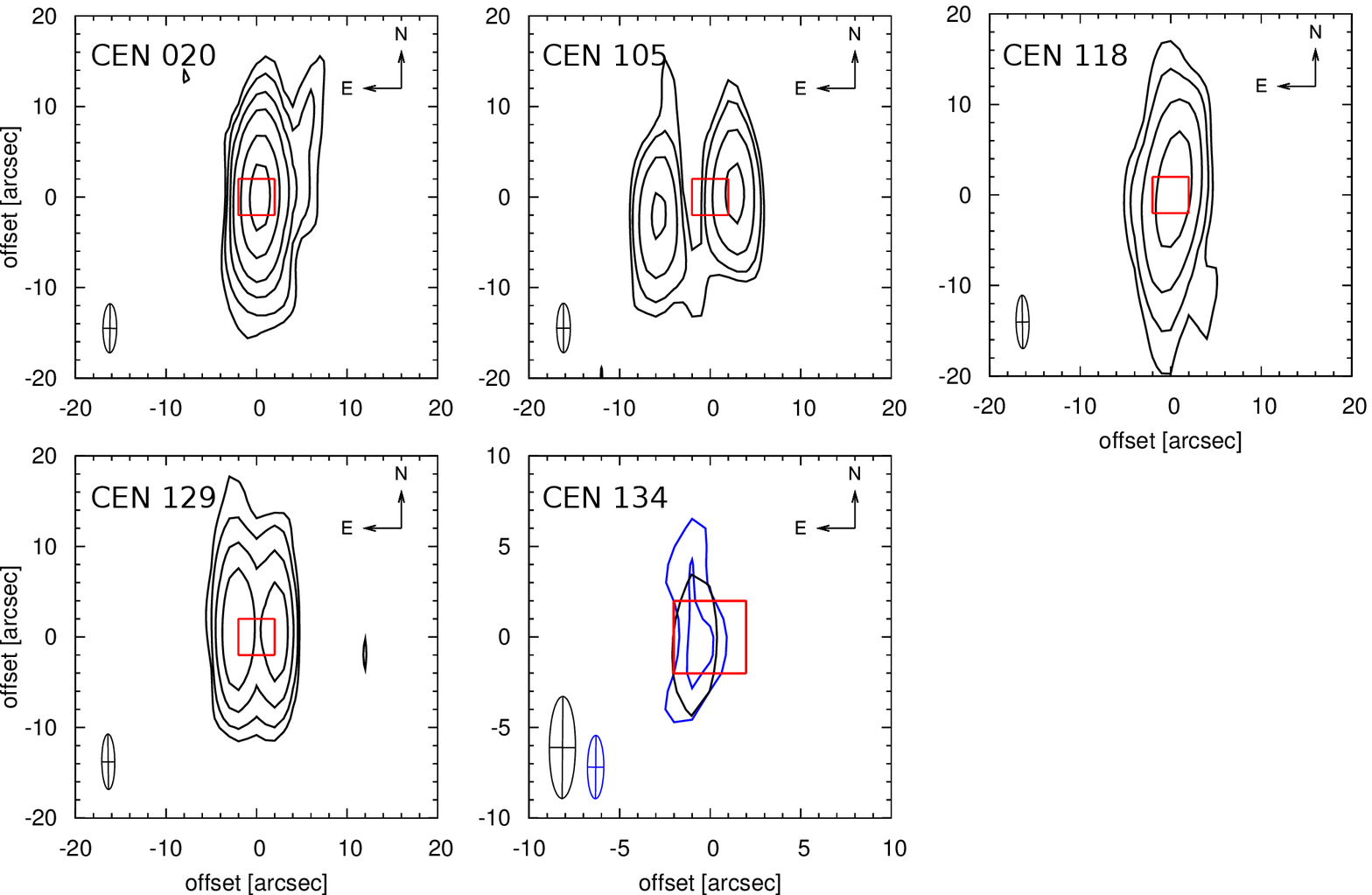}
\caption{Radio morphologies at 5.5~GHz of the MRCR-SUMSS sample. The
  size and orientation of the restored beam is given in the bottom
  left of each panel. The red box indicates the size and location of
  the SINFONI maps presented in Fig.~\ref{fig:spec1}. For
  NVSS~J144932$-$385657, given the short observation time of this
  source and hence the deformed beam, we also present its 4.8~GHz
  observations from \citet{Bryant2009a}, who observed a part of their
  sample at 4.8~GHz and 8.64~GHz.}
\label{fig:radioMorphologies}
\end{center}
\end{figure}

The radio morphologies are shown in Fig.~\ref{fig:radioMorphologies}.
They generally confirm those previously measured at 1.4~GHz and
2.4~GHz with larger beams. In NVSS~J002431$-$303330 we detect a
fainter second component to the southwest of the main radio emission
below the detection threshold of previous observations. In
NVSS~J234235$-$384526 we tentatively detect a radio core that is
coincident with the galaxy. Radio sizes are given in
Tab.~\ref{tab:atcaresults}. The largest angular scale (LAS) gives the
separation between the two lobes, if the source is resolved, or the
deconvolved size, if it is not.

\subsubsection{Relative astrometric alignment of the radio and SINFONI data}
\label{sss:ancillaryDataSetsRelativeAlignment}

Studying the effects of the radio jet on the interstellar medium of
high$-$redshift galaxies requires an accurate relative alignment between
the radio and near-infrared data sets to better than an arcsecond,
i.e., better than the absolute astrometry of the VLT. Unfortunately,
we did not detect the radio core of most of our galaxies with extended
radio lobes. Moreover, owing to the small field of view of SINFONI,
aligning our data cubes accurately within the World Coordinate System
(WCS) is not trivial. We therefore register our cubes relative to the
K-band imaging of \citet{Bryant2009a}, which is accurately aligned with
the WCS, and assume that the radio frame of ATCA aligns well with
the WCS, to better than 1\arcsec\ \citep[][]{Broderick2007}. For
compact radio sources (LAS~$\lesssim$~2.0\arcsec\ in
Tab.~\ref{tab:atcaresults}), we assume that the K-band
continuum is aligned with the radio source, corresponding to the
assumption that the radio emission in compact sources originates from
the nucleus of the galaxy. Figure~\ref{fig:radioMorphologies} shows the
radio contours of the MRCR-SUMSS and CENSORS sources, and the red box
gives the adopted position of the SINFONI maps based on this method.

\section{Results}
\label{s:PresentationOfOurAnalysisTools}

\begin{figure}
\centering
\includegraphics[width=0.8\textwidth]{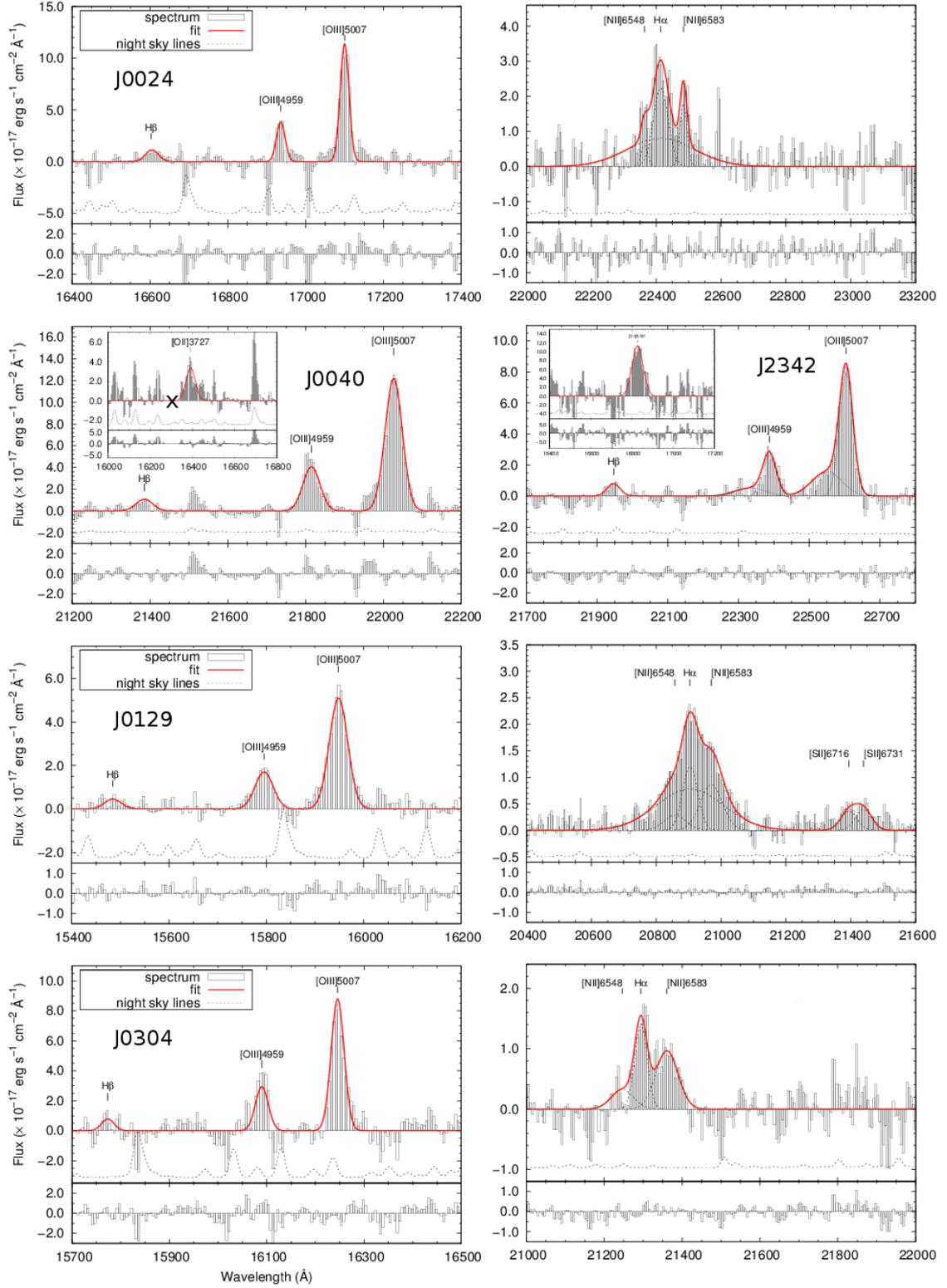}
\caption{Integrated spectra of our sources.  Gaussian fits to detected
  lines are plotted as solid red lines. Below each spectrum we show a
  typical night-sky spectrum to illustrate the position of bright
  night-sky lines (dashed lines). This spectrum is not to scale; the
  night-sky lines shown are brighter than the emission lines from our
  targets. The bottom panel shows the fit
  residuals.} \label{fig:spec1}
\end{figure}

\begin{figure}
\centering
\includegraphics[width=0.8\textwidth]{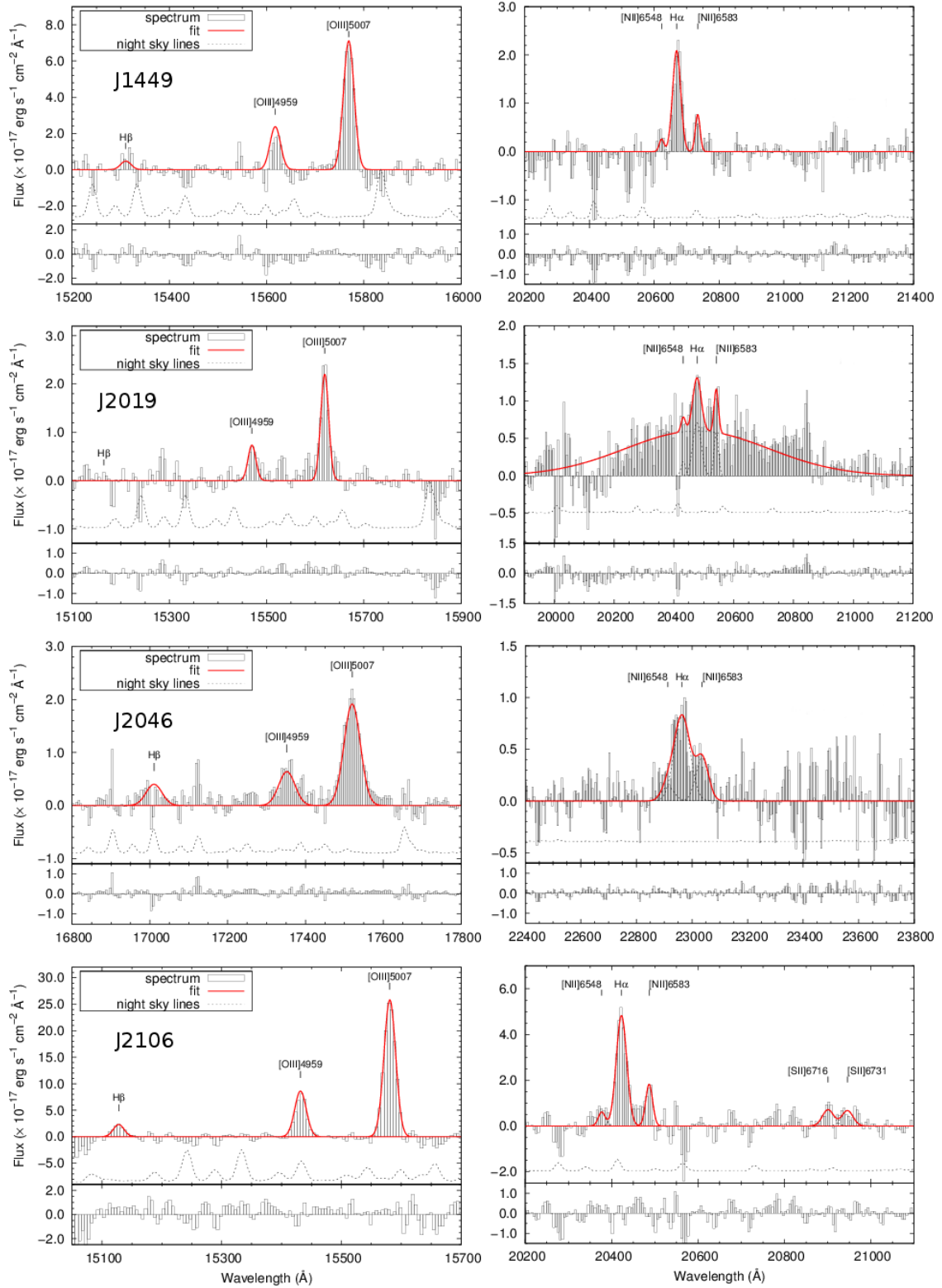}
\caption{Integrated spectra of our sources.  Gaussian fits to detected
  lines are plotted as solid red  lines. Below each spectrum we show a
  typical night-sky spectrum to illustrate the position of bright
  night-sky lines (dashed lines). This spectrum is not to scale; the
  night-sky lines shown are typically at least factors of a few
  brighter than the emission lines from our targets. The bottom panel
  shows the fit residuals.} \label{fig:spec2}
\end{figure}

\begin{figure}
\begin{center}
\includegraphics[width=0.8\textwidth]{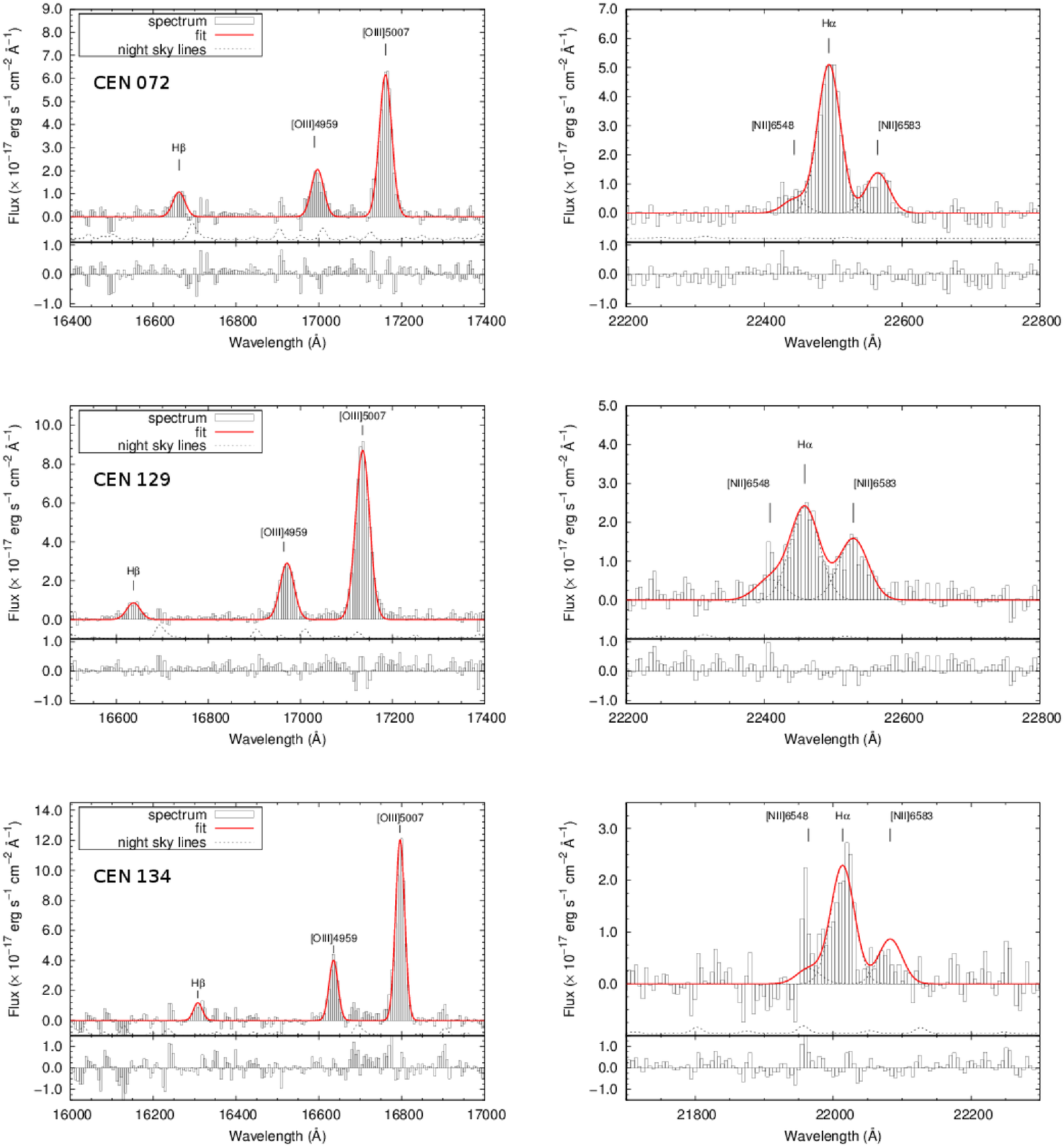}
\caption{Spectra of the CENSORS sample, integrated over pixels where
  the [OIII] emission line is detected at SNR$>$5. The left column shows
  spectra centered on the H$\beta$ and [OIII] doublet emission lines and the
  right column is centered on the H$\alpha$ + [NII] complex. All lines are
  well fitted by single Gaussians with the parameters listed in 
  Tab.~\ref{tab:spec1}. Below each spectrum we show a typical
    night-sky spectrum to illustrate the position of bright night-sky
    lines (dashed lines). This spectrum is not to scale; the night-sky
    lines shown are typically at least factors of a few brighter than
    the emission lines from our targets. The bottom panel shows the
    fit residuals.} \label{fig:spec3}
\end{center}

\end{figure}

For each galaxy we show integrated spectra and emission-line maps of
[OIII]$\lambda$5007 surface brightness, relative velocities, and FWHM
line widths (Figs.~\ref{fig:spec1}~to~\ref{fig:spec3} and
Tabs.~\ref{tab:spec1} and~\ref{tab:spec2}). Unless stated otherwise,
we give intrinsic FWHMs, $FWHM_{intrinsic}$, that are corrected for
instrumental resolution $FWHM_{inst}$, setting $FWHM_{intrinsic} =
\sqrt{FWHM_{obs}^{2}-FWHM_{inst}^{2}}$. The instrumental resolution,
$FWHM_{inst}$, is wavelength dependent and was measured from the width
of night-sky lines. Maps are only given for spatial pixels where the
signal-to-noise ratio of the line core exceeds 5. We used a Monte
Carlo method to confirm that this was a good value with which to robustly measure
the line properties in spite of strong non-Gaussianities in the noise
related to the imperfect night-sky line subtraction, bad pixels, and
potentially intrinsic line profiles.

Integrated spectra include all pixels where [OIII]$\lambda$5007 is
detected at a significant level. We adopt the redshift estimated from
the brightest pixels near the center of the galaxy as systemic. Before
adding the spectrum of a pixel, we shift it to the systemic redshift
in order to avoid artificial broadening of the line in the integrated
spectrum by the large-scale velocity gradient.

\begin{figure}
\centering
\includegraphics[width=0.8\textwidth]{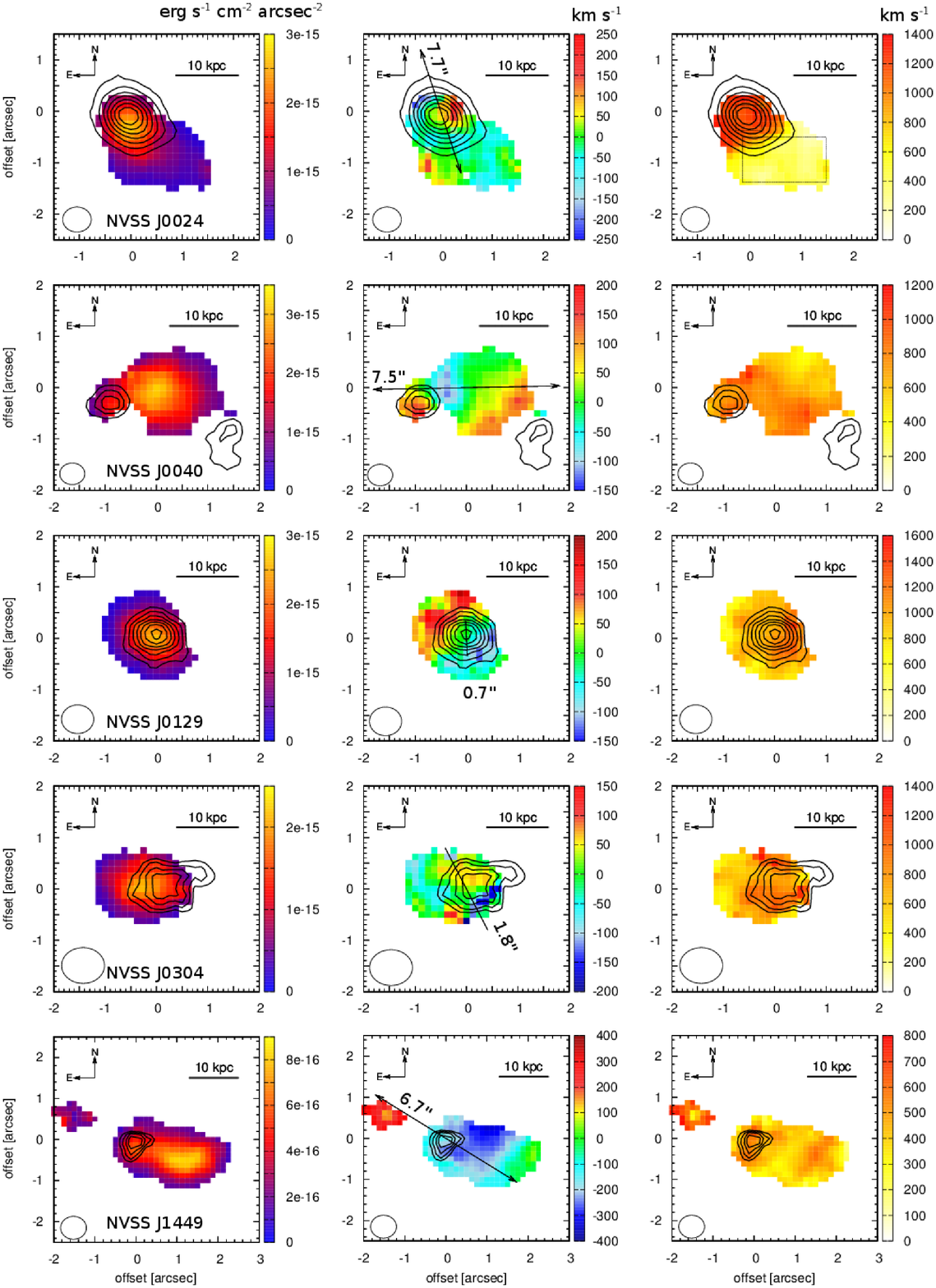}
\caption{Maps of [OIII] surface brightness ({\it left}), velocity
  ({\it center}), and FWHM line width ({\it right}) of our nine sources
  (from {\it top} to {\it bottom}). All maps are
  4\arcsec$\times$4\arcsec wide, except for NVSS~J144932$-$385657 and
  NVSS~J210626$-$314003, where they are 5\arcsec$\times$5\arcsec
  wide. The circle in the bottom left represents the FWHM size of the
  seeing disk. Contours mark the continuum where detected. Continuum
  levels begin at 3$\sigma$ and then increase in steps of 1$\sigma$.
  The solid black line in the velocity maps indicates the axis of the
  radio emission from our ATCA data, or from \citet{Broderick2007} if
  the source is compact. Numbers give half the largest angular size in
  arcsec.
\label{fig:maps1}}
\end{figure}

For each galaxy we also mapped the surface brightness, relative
velocity to the systemic redshift and the FWHM line widths of
[OIII]$\lambda$5007 (Fig.~\ref{fig:maps1} to~\ref{fig:maps3}) by
fitting Gaussian profiles to the lines extracted from small apertures
across the cube. Aperture sizes are 3~pixels $\times$ 3~pixels,
corresponding to 0.4\arcsec\ $\times$ 0.4\arcsec, or 5~pixels $\times$
5~pixels (0.6\arcsec\ $\times$ 0.6\arcsec) for the faintest regions of
the source. This helps to improve the signal-to-noise ratio of the
data, but still oversamples the seeing disk and avoids loss of spatial
information. Since the sizes of the extended emission-line
regions, $S_{maj,obs}$, are typically only a few times larger than
the size of the seeing disk, we list sizes of semi-major and
semi-minor axes that are corrected for the broadening of the PSF
$S_{PSF}$, $S_{maj,intrinsic}$, by setting $S_{maj,intrinsic} =
\sqrt{ S_{maj,obs}^2-S_{PSF}^2}$ along the same position angle. The
same method was applied to the size along the semi-minor axis, where
resolved.

Contours in Fig.~\ref{fig:maps1} show the line-free continuum emission
for the galaxies where the continuum was detected. In most galaxies
the continuum is only detected after collapsing the line-free cube
along the wavelength axis.  However, we detect relatively bright
continuum emission in NVSS~J002431$-$303330 and NVSS~J201943$-$364542,
which we need to subtract from the spectra before fitting the emission
lines. To perform this subtraction, we mask strong emission lines and
strong night-sky line residuals and fit a fifth-order polynomial over
their whole spectrum, which we subtract afterward.

\subsection{Description of individual sources}
\label{ss:NVSSssourcesDescription}

\subsubsection{NVSS~J002431$-$303330}
\label{sss:J0024}

NVSS~J002431$-$303330 is a double radio source, dominated by a bright
component associated with the optical counterpart, and a weaker
component at 16\arcsec\ toward southwest
(Fig.~\ref{fig:radioMorphologies}). We find this source at a redshift
of $z=2.415 \pm 0.001$, which is in good agreement with the estimate
of Johnston et al. (in prep.) from the rest-frame UV lines, with
$z_{\mathrm{\textsc{uv}}}=2.416 \pm 0.001$. In the H band, we detect
the [OIII]$\lambda 4959,5007$ doublet and H$\beta$. In the K band, we
find H$\alpha$ and [NII]$\lambda$6548,6583, which are strongly blended
owing to their large intrinsic widths. The H$\alpha$ line has a broad component
with FWHM~=~3250~km s$^{-1}$. Figure~\ref{fig:spec1} shows the
integrated spectrum of this source. All line properties are listed in
Tab.~\ref{tab:spec1}.

As shown in Fig.~\ref{fig:maps1}, NVSS~J002431$-$303330 has a strong
continuum associated with a bright emission-line region with
[OIII]$\lambda$5007 surface brightness of $(5-25)\times 10^{-16}$ erg
s$^{-1}$ cm$^{-2}$ arcsec$^{-2}$. Line widths in this region are very
broad, FWHM $\sim$ 1200~km s$^{-1}$, and the velocity field is
perturbed with two small, unresolved regions that show abrupt velocity
jumps relative to their surroundings with relative redshifts of about
250 km s$^{-1}$ in each region. This area extends over
$\sim$~1.0\arcsec$\times$1.0\arcsec\ around the peak of the continuum
and [OIII]$\lambda$5007 line emission, corresponding to 8~kpc at
$z=2.415$. The H$\alpha$ surface brightness in this region is
$\Sigma_{\mathrm{H}\alpha} \sim (5 - 9) \times 10^{-16}$ erg s$^{-1}$
cm$^{-2}$ arcsec$^{-2}$.

Toward the southwest, the line emission becomes fainter, but can be
traced out over another 2\arcsec, with a typical [OIII]$\lambda$5007
surface brightness of $\Sigma_{\mathrm{[OIII]}}=(1-5)\times 10^{-16}$
erg s$^{-1}$ cm$^{-2}$ arcsec$^{-2}$. The gas is more quiescent, with
line widths of FWHM$=$300$-$400 km s$^{-1}$. We show integrated
spectra of both regions in Fig.~\ref{fig:J0024_OIIISpectra}. The box
in the right panel of Fig.~\ref{fig:maps1} shows the region from which
we extracted the narrow-line emission.  This extended emission-line
region to the southwest extends along the axis between the two radio
lobes.

\subsubsection{NVSS~J004000$-$303333}
\label{sss:J0040}

NVSS~J004000$-$303333 is a double radio source with a size of
17\arcsec\ (Fig.~\ref{fig:radioMorphologies}). With SINFONI we find
the [OIII]$\lambda\lambda$4959,5007 doublet and H$\beta$ in the K band
at wavelengths that correspond to $z = 3.399 \pm 0.001$. The H$\alpha$
and [NII]$\lambda\lambda6548,6583$ lines fall outside the atmospheric
windows. In the H band, we detect the [OII]$\lambda$3727 doublet. The
two lines of the doublet are too close to each other to be spectrally
resolved with our data (Fig.~\ref{fig:spec1}). All line properties are
listed in Tab.~\ref{tab:spec1}.

Line emission extends over 2.5\arcsec$\times$1.5\arcsec\ with surface
brightnesses between $\Sigma_{\mathrm{[\textsc{Oiii}]}} \sim 5 \times
10^{-16}$ erg~s$^{-1}$ cm$^{-2}$ arcsec$^{-2}$s and $3 \times
10^{-15}$ erg s$^{-1}$ cm$^{-2}$ arcsec$^{-2}$
(Fig.~\ref{fig:maps1}). Faint continuum emission is found associated
with a knot to the very east of the emission-line region, and about
1.5\arcsec\ southwest from the center,  outside the bright line
emission. The K-band image of \citet{Bryant2009a} shows a very similar
continuum morphology, likewise at low signal-to-noise ratio.

The local velocities of [OIII]$\lambda$5007 fall monotonically from the
southwest to the northeast with a total gradient of about 300 km
s$^{-1}$. The knot in the far east shows an abrupt velocity increase
of 300 km s$^{-1}$ relative to the nearby blueshifted gas. The
line widths are lower in the north (FWHM~=~200$-$400~km
s$^{-1}$) than in the south (FWHM~=~700$-$1000~km s$^{-1}$).

At fainter flux levels than shown in Fig.~\ref{fig:maps1}, but still
above 3$\sigma$, we detect another source of line emission at a
distance of about 2\arcsec\ to the south from the radio galaxy (about
15~kpc at z~$\sim$~3). The redshift of this second source is
$z_{\mathrm{south}} = 3.395 \pm 0.001$, i.e., it is blueshifted by
$350 \pm 90$~km s$^{-1}$ relative to the radio galaxy. This source is
shown in Fig.~\ref{fig:J0040_SN3} and is  discussed in
\S\ref{s:additionalLineEmittersTracersOfEnvironmentDynamicalProbes}.

\subsubsection{NVSS~J012932$-$385433}
\label{sss:J0129}

The SINFONI maps of NVSS~J012932$-$385433 are shown in
Fig.~\ref{fig:maps1}. We find the optical emission lines at  redshift
$z = 2.185 \pm 0.001$. The radio source is compact with a deconvolved
size of 0.7\arcsec, and is associated with the optical counterpart. 

[OIII]$\lambda\lambda$4959,5007 is bright in the H band.  H$\beta$ is
detected at 5.6$\sigma$. In the K band, H$\alpha$ and
[NII]$\lambda\lambda$6548,6583 are detected and strongly
blended. H$\alpha$ also shows a broad component with FWHM $\sim$
3500~km s$^{-1}$. The $[$SII$]$ doublet is clearly detected. The two
components are also strongly blended owing to their intrinsic width
(Fig.~\ref{fig:spec1}).  All line properties are listed in
Tab.~\ref{tab:spec1}.

The emission-line region is extended over
1.6\arcsec$\times$1.2\arcsec. It is brighter in the center with
$\Sigma_{\mathrm{H}\alpha} \sim (1.0 - 1.7) \times 10^{-15}$
erg~s$^{-1}$ cm$^{-2}$ arcsec$^{-2}$ and fades toward the
periphery. We detect continuum emission coincident with the emitting
gas. The [OIII]$\lambda$4959,5007 lines show a clear, relatively
regular velocity gradient of $\Delta$v~$\sim$~350~km s$^{-1}$ along a
northeast-southwest axis. The lines are more narrow toward
the southeast, with FWHM $\sim$ 700$-$800~km s$^{-1}$. In the
northwest FWHMs are higher, $\sim$ 900$-$1000~km s$^{-1}$.

\subsubsection{NVSS~J030431$-$315308}
\label{sss:J0304}

NVSS~J030431$-$315308 is a single, relatively compact source at 9~GHz
and 5.5~GHz, with a deconvolved size of 1.8\arcsec\ in our
highest resolution data at 9~GHz (Fig.~\ref{fig:radioMorphologies}).
The [OIII]$\lambda\lambda$4959,5007 doublet is clearly detected in the
H band with SINFONI and is well fitted with single Gaussians
(Fig~\ref{fig:spec1}). The same holds for the H$\alpha$ and
[NII]$\lambda\lambda$6583 lines. The $[$SII$]$ doublet is not
detected. All line properties are listed in Tab.~\ref{tab:spec1}.

The line emission is marginally
spatially resolved with a size of 1.5\arcsec$\times$1.5\arcsec, and a
PSF with FWHM=1.2\arcsec$\times$1.0\arcsec, the largest in this
program (Tab.~\ref{tab:obslog}). Faint continuum emission is also
detected, at a slightly different position ($\sim$~0.5\arcsec\ to the
west) from the peak in [OIII]$\lambda$5007 surface brightness, but at
the same position as the peak of H$\alpha$ surface brightness.
The velocity maps show two small redshifted (by 50$-$100 km s$^{-1}$)
regions north and south of the continuum, and uniform velocities in
the rest of the source. Line widths are between 500 and 1200 km
s$^{-1}$ and higher in the western parts of the emission-line region
associated with the continuum. 

\subsubsection{NVSS~J144932$-$385657}
\label{sss:J1449}

NVSS~J144932$-$385657 has one of the largest radio sources in our
sample; the lobes are offset by 7.5\arcsec\ relative to each other. We
find the [OIII] line at $z = 2.149 \pm 0.001$. In the H band, we
detect the [OIII]$\lambda\lambda$4959,5007 doublet and H$\beta$
(Fig.~\ref{fig:spec2}). In the K band, H$\alpha$ and
[NII]$\lambda$6583 are narrow enough not to be blended. FWHM$=$350~km
s$^{-1}$ for H$\alpha$, and the width of [NII]$\lambda$6583 is
dominated by the spectral resolution.

NVSS~J144932$-$385657 has a very extended emission-line region of nearly
4\arcsec\ ($\sim 30$ kpc at $z \sim 2$) along a northeast-southwest axis (Fig.~\ref{fig:maps1}). We identify two parts: a
large, elongated region, which coincides with the continuum and
extends over another 2\arcsec\ toward the southwest, and a fainter,
smaller region in the northwest. Surface brightnesses of
[OIII]$\lambda$5007 are between $1\times 10^{-16}$ erg s$^{-1}$
cm$^{-2}$ arcsec$^{-2}$ and $10\times 10^{-16}$ erg s$^{-1}$
cm$^{-2}$ arcsec$^{-2}$. The northwestern region is near the edge of
the SINFONI data cube, and it is therefore possible that it extends
farther beyond the field of view of our data. Both emission-line
regions are aligned with the axis of the radio jet.

The velocity offset between the two regions is about 800 km s$^{-1}$.
While the compact northeastern part has a uniform velocity field with
a redshift of about v $\sim 400$~km s$^{-1}$, the kinematics in the
very extended southwestern region are more complex with a maximum
blueshift of about $-400$ km s$^{-1}$, before the velocities
approach the systemic redshift again at the largest radii. Line
widths are FWHM=200$-$500 km s$^{-1}$ in the southwest, and up to 800
km s$^{-1}$ in the northeast. In the southwestern region we find
elevated widths in particular near the continuum and at about
1.5\arcsec, a  distance associated with the sudden velocity shift from
$-400$ km s$^{-1}$ to the systemic velocity.
 
In Fig.~\ref{fig:J1449}, we compare the surface brightness maps of
H$\alpha$ and [NII]$\lambda$6583, obtained by fitting three Gaussians
corresponding to H$\alpha$, [NII]$\lambda$6548, and [NII]$\lambda$6583
with the velocities and line widths measured from [OIII]$\lambda$5007,
and leaving the line flux as a free parameter. The map highlights the
similarity of the H$\alpha$ and [NII]$\lambda$6583 morphologies, which
justifies our assumption that the Balmer and the optical forbidden
lines originate from the same gas, at least at the spatial resolution
of our data.

\subsubsection{NVSS~J201943$-$364542}
\label{sss:J2019}

The radio source of NVSS~J201943$-$364542 is very extended, with LAS =
14.7\arcsec, corresponding to $\sim$ 120~kpc at $z = 2.1$
(Fig.~\ref{fig:radioMorphologies}). In the H band with SINFONI, we
detect the [OIII]$\lambda\lambda$4959,5007 doublet at a redshift of $z
= 2.120 \pm 0.001$, but not H$\beta$. In the K band, in addition to
the narrow components of H$\alpha$ and [NII]$\lambda$6583, we find a
broad H$\alpha$ emission line (FWHM $\ge$ 8000~km s$^{-1}$) from a
compact region aligned with the nucleus. We  come back to this
line in \S\ref{s:additionalLineEmittersTracersOfEnvironmentDynamicalProbes}.
Integrated line properties are listed in Tab.~\ref{tab:spec1}. 

NVSS~J201943$-$364542 has a compact emission-line region of
1.0\arcsec\ with an [OIII]$\lambda$5007 surface brightness
$\Sigma_{\mathrm{[\textsc{Oiii}]}} = (0.5 - 1.0) \times 10^{-15}$
erg~s$^{-1}$ cm$^{-2}$ arcsec$^{-2}$ (Fig.~\ref{fig:maps2}). It also has
 relatively bright continuum emission associated with the line
emission. The velocity field of the narrow-line component shows a
scatter of $\le$ 200 km s$^{-1}$, with the highest velocities reached
in the far east and west. The line widths are up to 800 km s$^{-1}$.

At about 3\arcsec\ to the southeast we marginally detect another compact
line emitter at a very similar redshift $z_{\mathrm{south}} = 2.116
\pm 0.001$ (Fig.~\ref{fig:J2019_Spectrum2ndRegion}). The proximity on
the sky and in redshift suggests that both sources are physically
related with the radio galaxy. We will discuss this source in more detail in
\S\ref{s:additionalLineEmittersTracersOfEnvironmentDynamicalProbes}.

\subsubsection{NVSS~J204601$-$335656}
\label{sss:J2046}

NVSS~J204601$-$335656 has a compact radio source with a deconvolved
size of 1.8\arcsec\ in our data, consistent with the
1.6\arcsec\ previously found by \citet{Broderick2007}. With SINFONI in
the H band, we detect the [OIII]$\lambda\lambda$4959,5007 doublet and
H$\beta$ at $z = 2.499 \pm 0.001$. In the K band, H$\alpha$ and the
[NII]$\lambda\lambda$6548,6583 doublet are blended but well fitted
with a single Gaussian component for each line
(Fig.~\ref{fig:spec2}). The $[$SII$]\lambda\lambda$6716,6731 doublet
is not detected.  Table~\ref{tab:spec1} summarizes the line properties
of NVSS~J204601$-$335656.

The line emission is marginally spatially resolved with a FWHM size of
1\arcsec\ compared to a 0.7\arcsec$\times$0.6\arcsec\ PSF.  The
emission-line region is associated with a faint continuum source; the region is roughly circular and extends over 1.0\arcsec, with
$\Sigma_{\mathrm{[OIII]}} = (0.3 - 1.7) \times 10^{-15}$ erg~s$^{-1}$
cm$^{-2}$ arcsec$^{-2}$.

NVSS~J204601$-$335656 has a small velocity offset of 150 km s$^{-1}$
which rises from northeast to southwest. Typical line widths are
FWHM$=900-1200$~km s$^{-1}$.

\subsubsection{NVSS~J234235$-$384526}
\label{sss:J2342}

NVSS~J234235$-$384526 is an extended radio source with two lobes at a
relative distance of
9.8\arcsec\ \citep[Fig.~\ref{fig:radioMorphologies}, see
  also][]{Broderick2007}. It is at  redshift $z = 3.515 \pm 0.001$,
where the [OIII]$\lambda\lambda$4959,5007 doublet and the H$\beta$
emission lines fall into the K band. Fitting the
[OIII]$\lambda\lambda$4959,5007 line profiles adequately requires two
components per line (Fig.~\ref{fig:spec1}), a narrow component with
FWHM = 360~km s$^{-1}$ (which we consider to be at the systemic
redshift), and a broader blue wing with FWHM = 1300~km s$^{-1}$,
blueshifted by 700~km s$^{-1}$. In the H band, we detect the
[OII]$\lambda$3727 emission lines. Line properties are listed in
Tab.~\ref{tab:spec1}.

Line emission extends over 2.0\arcsec$\times$1.0\arcsec\ along an axis
going from northeast to southwest, with surface brightness
$\Sigma_{\mathrm{[\textsc{Oiii}]}} = (5 - 30) \times
10^{-16}$~erg~s$^{-1}$ cm$^{-2}$ arcsec$^{-2}$
(Fig.~\ref{fig:maps2}). The velocity map shows a gradient of about
400~km s$^{-1}$ rising from the northeast to the southwest, and well
aligned with the axis of the radio source.  Line widths are larger in
the center of the galaxy (FWHM $\simeq 1000-1200$~km s$^{-1}$) than in
the periphery (FWHM $\simeq$ 200-400~km s$^{-1}$). We do not detect
any continuum emission.

\subsection{CENSORS sample}
\label{s:CENSORS}

\subsubsection{NVSS~J094949$-$211508 (CEN~072)}

This source has a redshift of z~=~$2.427 \pm 0.001$, in agreement with
the previous UV redshift measured by
\citet{Brookes2008CENSORS}. Because there was a coordinate mismatch,
we do not have new radio measurements at 5.5 and 9.0~GHz for this
source. \citet{Best2003CENSORS} found LAS$<$0.7\arcsec, which is small
enough to infer that the extended emission-line region is larger than
the radio emission.

The integrated spectrum shown in Fig.~\ref{fig:spec3} shows that
[OIII]$\lambda\lambda$4959,5007 and H$\beta$ are well detected in the
H band, and H$\alpha$ and [NII]$\lambda\lambda$6548,6583 in the
K band, The lines are not very strongly
blended. [SII]$\lambda\lambda$6716,6731 and [OI]$\lambda$6300 are not
detected.

Line emission extends over
1.7\arcsec$\times$1.2\arcsec\ (14~kpc$\times$10~kpc) along a
northwest-southeast axis (Fig.~\ref{fig:maps3}). It has a
velocity gradient of $\sim$~300~km s$^{-1}$.
The stellar continuum is associated with the northwestern
part of the emission-line region, where the line widths also
reach their maximum (FWHM~$\sim$~800~km s$^{-1}$), compared to
FWHM~$\sim$~300-500~km s$^{-1}$ in the southeast.

\subsubsection{NVSS~J095226$-$200105 (CEN~129)}

This source is at  redshift z~=~$2.422 \pm 0.001$, in agreement with
the previous estimate of z$=$2.421 based on rest-frame UV lines
\citep{Brookes2008CENSORS}. The radio morphology is resolved in our
ATCA observation at 5.5 and 9.0~GHz (see
Fig.~\ref{fig:radioMorphologies}), but was compact in the previous
1.4~GHz data of \citet{Best2003CENSORS}. We find two radio lobes along
an east-west axis along a position angle of 95$^\circ$, separated by
2.5\arcsec (20~kpc at z$=$2.4). 

The integrated spectrum of CEN~129 shows H$\beta$ and the
[OIII]$\lambda\lambda$4959,5007 doublet in the H band, and H$\alpha$
and the [NII]$\lambda\lambda$6548,6883 lines in the K band
(Fig.~\ref{fig:spec3} and Tab.~\ref{tab:spec2}).  The emission-line
morphology of CEN~129 is fairly spherical, with a small extension
toward the northwest, and a size of
2.3\arcsec$\times$1.7\arcsec\ (19~kpc$\times$14~kpc). The major axis
is along the northwest-southeast direction. The stellar continuum
is detected at the center of the line emission. The velocity field has
a gradient of $\sim$~200~km s$^{-1}$, and is roughly aligned with the
radio axis, along an east-west axis. The northwestern extension has
the most blueshifted emission ($-$200~km s$^{-1}$). Line widths are
fairly uniform with FWHM~$\sim$~600-700~km s$^{-1}$.

\subsubsection{NVSS~J094949$-$213432 (CEN~134)}

The integrated spectrum of CEN~134 shows H$\beta$ and the
[OIII]$\lambda\lambda$4959,5007 doublet in the H band, and H$\alpha$
and [NII]$\lambda$6583 in the K band. We find a redshift of z~=~$2.355
\pm 0.001$ for CEN~134, in good agreement with the value estimated from
rest-frame UV lines \citep{Brookes2008CENSORS}. At 1.4~GHz, the radio
source is very extended, with LAS=22.4\arcsec\ along a position angle
of 125$^\circ$ \citep[][]{Best2003CENSORS}. At 5.5~GHz we measure
LAS~=~21.9\arcsec\ and PA~=~131$^\circ$, but we do not detect the
second lobe in the 9.0~GHz observations.

Line emission in CEN~134 extends over
3.1\arcsec$\times$1.8\arcsec\ (25~kpc$\times$15~kpc), with the major
axis going from the northwest to the southeast
(Fig.~\ref{fig:maps3}). Continuum emission is detected in the southern
part of the emission-line region. The velocity field shows two
blueshifted regions, one south of the continuum, one at the very
north of the emission-line region, with velocities of about $-120$ km
s$^{-1}$ relative to the velocities near the center. Line widths are
higher in the south near the continuum position, with
FWHM~$\sim$~600~km s$^{-1}$. In the north, the gas is more quiescent
with FWHM$\sim$~300~km s$^{-1}$.

\section{Ensemble properties of the CENSORS and MRCR-SUMSS samples}
\label{sec:ensembleproperties}
After discussing each of our targets individually and in detail, we
 now turn to characterizing the overall properties of our sample.
Although we do not have a complete sample in a strict statistical
sense, this is a fairly common situation for studies of the spatially
resolved properties of small to mid-sized samples of high-redshift
galaxies \citep[e.g.,][]{nmfs06, nmfs09}, and our sample size is
comparable to most of these studies. It is also important to note that
our study is a parameter study, not a population study. This means that we
wish to analyze certain source properties as a function of the AGN
characteristics of our sources.  Therefore, we have to sample the
range of AGN properties as uniformly as we can, and can put less
emphasis on matching, e.g., the shape of the radio luminosity
function. For this reason, we do not require a statistically complete
sample in order to identify global trends in our data.

\subsection{Rest-frame optical continuum} 

We detect rest-frame optical continuum emission in 12 sources, 9 from
the MRCR-SUMSS, and 3 from the CENSORS sample. At redshifts $z=2-3$, the
observed H and K bands correspond roughly to the rest-frame V and
R bands, and at redshifts $z=3-4$ to the rest-frame B and V bands.

Since continuum fluxes in all sources are too faint to measure the
detailed spectral profiles or even spectral slopes, we merely extract
the spectrally integrated continuum image (Figs.~\ref{fig:maps1}
to~\ref{fig:maps3}).  We generally find only one spatially
unresolved continuum source per target at most, with one notable
exception. In NVSS~J004000$-$303333 we find two very faint unresolved
continuum emitters of about equal flux, perhaps indicating an ongoing
interaction of two galaxies, or AGN light scattered on extended dust
\citep[e.g.,][]{hatch09}. Both blobs are roughly along the radio jet
axis, somewhat reminiscent of the alignment effect found in more
powerful HzRGs \citep[][]{tadhunter98,vernet01}. In
NVSS~J234235$-$384526 we do not detect the continuum at all.

\subsection{Morphology of the extended emission-line gas}

Gas morphologies and kinematics are the two primary sets of
constraints where the advantages of imaging spectrographs become
particularly evident. Among our 12 targets, 10 are spatially
resolved into a maximum of 10 elements along the
major axis of the bright emission-line regions.

The emission-line morphologies and kinematics of our sources are very
diverse. Isophotal sizes down to the 3$\sigma$ detection limit of a few
$10^{-17}$ erg s$^{-1}$ cm$^{-2}$ arcsec$^{-1}$ range from about
5$-$6~kpc, corresponding to our resolution limit, to very extended
sources where line emission is detected over at least 30~kpc (e.g., in
NVSS~J144932$-$385657), and potentially more because we cannot exclude
in all cases (e.g., NVSS~J144932$-$385657) that parts of the
emission-line region fall outside the 8\arcsec$\times$8\arcsec\ field
of view of SINFONI.  Resolved emission-line regions are often
elongated with ellipticities between 0.2 and 0.7. We do find a
correlation between elongation and size, but cannot rule out that this
is an artifact from the relatively low spatial resolution.  Typical
emission-line surface brightnesses are between a few $10^{-16}$ erg
s$^{-1}$ cm$^{-2}$ arcsec$^{-2}$ and a few $10^{-15}$ erg s$^{-1}$
cm$^{-2}$ arcsec$^{-2}$ (Figs.~\ref{fig:maps1} to \ref{fig:maps3}).

Eight of our galaxies have spatially-resolved gas and a single
rest-frame optical continuum peak. In those cases we can evaluate
whether the line emission extends symmetrically about the nucleus
based on the continuum peak being an approximation for the location of
the central regions of the HzRGs \citep[see][for a
  justification]{Nesvadba2008}.  It is interesting that this is not
always the case; in particular, we note a trend that the galaxies with
the most regular velocity fields also appear to have gas that is well
centered about the nucleus, as would be expected from a more
dynamically relaxed system. While lopsidedness does exist even in
isolated low-z galaxies, it would need to reach extreme levels to be
seen at a resolution of 5 kpc and more. In NVSS~J004000$-$303333, the
source with the two equally faint continuum emitters, most of the gas
is between the two continuum sources. This may somewhat favor the
interpretation that the continuum is from two regions of scattered
light rather than a merging galaxy pair, in which case we would expect at
least some line emission to be associated with the nuclear regions of
each interacting galaxy.

\subsection{Blending of circumnuclear and extended emission}
\label{ssec:toymodels}

A potential worry for our morphological study is that unresolved
circumnuclear emisson might be bright enough to dominate the
emission-line maps even beyond the central PSF.  Even very compact
($\le 1$~kpc) emission-line regions akin to classical narrow-line
regions (NLRs), if bright enough, could bias our measurements. 
In radio galaxies, where the direct view into
the circumnuclear regions is obscured, this may be less important than
in quasars, but since some of our sources do show a prominent nucleus,
a more quantitative analysis is in order. 

We constructed a suite of toy data cubes with and without a bright NLR
embedded in extended, fainter gas with more quiescent kinematics.  The
NLRs were approximated by a bright point source, the extended line
emission by a region of uniform surface brightness, with sizes,
spatial and spectral sampling, and a spatial resolution comparable to
our data (Fig.~\ref{fig:toymodel}). Both components have Gaussian line
profiles with FWHM$=$300 km s$^{-1}$ in the extended emission and
FWHM$=$800 km s$^{-1}$ for the circumnuclear gas. These widths were
derived from the final fitted data cube. The extended component has a
velocity gradient of 400~km s$^{-1}$ over the full source diameter. The
signal-to-noise ratios in the final data cubes are comparable to our
data, and for simplicity we assumed Gaussian noise. The beam
smearing due to the seeing was approximated by convolving the data
cube with a two-dimensional Gaussian in each wavelength plane.

For cubes with a dynamic range of at least 10 in gas surface
brightness between nuclear point source and extended emission in the
fitted data, we find an apparent increase in line width of
30$-$50\% at a distance of 1$\times$ the PSF from the nucleus
due to the NLR, and of 10$-$20\% at 1.5$\times$ that radius
(Fig.~\ref{fig:toymodel}); 10$-$20\% corresponds to our measurement
uncertainties. The NLR component in these galaxies is easily seen in
the integrated spectrum (Fig.~\ref{fig:toymodelspec}).

However, only one galaxy (NVSS~J144932$-$385657) has a dynamic range
as high as 10. In most cases, the nucleus is only about 4-6 times
brighter than the faintest extended emission (Fig.~\ref{fig:maps1} to
\ref{fig:maps3}). Such a contrast is not sufficient to produce the
observed strong increase in FWHM from 300 to 800~km s$^{-1}$. The
integrated line flux scales linearly with line core and FWHM. Because
of the line broadening, an increase in surface brightness of a factor
of 2.5-3 is therefore already implied by the line broadening, and the
core of the broader component does not make a large contribution to
the core of the measured line profile in the combined spectrum of
circumnuclear and extended component. At SNR~5$-$10, the line wings of
the circumnuclear component are hidden in the noise.

The broadening is stronger if the extended line emission is
distributed asymmetrically about the nucleus, as is the case
for CEN-134. For example, if we truncate the extended emission prior
to the smoothing with the seeing disk at 1 times the FWHM of the
PSF from the nucleus on one side, the central line width increases
by a factor of~2 compared to the symmetric case. However, the
circumnuclear spectral component then becomes  even more prominent in
the integrated spectrum. For an extended component that extends out
to at least 1.5 times the FWHM size of the PSF, the difference
 very rapidly becomes indistinguishable.

This shows that our objects are not comparable to low-redshift Seyfert
galaxies, for example, where a systemic and a narrow-line region,
often with very different properties, can clearly be
distinguished. The gas producing the relatively broad emission lines
must extend over larger radii, even if these regions are not clearly
resolved at the 5$-$8~kpc resolution of our data.

\subsection{Kinematics}
\label{ssec:kinematics}
Our sample shows a wide variety of kinematic properties. Velocity
fields range from very regular-- dominated by a single, smooth,
large-scale velocity gradient-- to very irregular.  The total velocity
gradients are typically on the order of v~$=$~200$-$300 km s$^{-1}$. Given
that the spatial resolution of our data is not very high for many of
our sources, this gradient may appear lower than the intrinsic
velocity gradient owing to beam-smearing effects. We used a set of
Monte Carlo simulations to estimate that beam-smearing may lower the
measured velocity gradients by about a factor of~2, comparable to
inclination (Collet et al. 2014, PhD~thesis). Overall, the line widths in our
galaxies are  relatively broad, FWHM$=$400$-$1000 km s$^{-1}$,
down to the spatial resolution limit of our data. This is more than
 can be attributed to the overall velocity shifts and beam
smearing.

Visual inspection of Figures~\ref{fig:maps1}~to \ref{fig:maps3} shows
that at most two galaxies, NVSS~J012932$-$385433 and CEN~072, have
regular velocity gradients without very obvious perturbations. All the
other galaxies have significantly perturbed velocity fields. Even the
very regular galaxy NVSS~J012932$-$385433 may have slight
irregularities in the blueshifted gas in a small region of the
southwestern hemisphere, which is strongly blurred by the size of the
seeing disk (Fig.~\ref{fig:maps1}). However, both galaxies have
irregular distributions of line widths, which are at variance with
quiescent disk rotation.

A good example for a galaxy with an irregular velocity field is
CEN~134 (Fig.~\ref{fig:maps3}). This galaxy, and others with
irregular velocity fields, exhibits sudden velocity jumps at least in
parts of the emission-line region, and at signal-to-noise levels where
this irregularity must be intrinsic to our sources over regions
consistent with at least the size of the seeing disk (which is
oversampled in the seeing-limited SINFONI mode, with 4$-$6 pixels per
FWHM of the PSF). This would not be the case for simple noise
features. It is interesting that these velocity jumps often coincide 
with the position of the continuum where we would expect the galaxy
nucleus, and hence the AGN. They are also associated with a local
broadening of the emission lines, but are not large enough to be the
sole cause of this broadening. We will come back to the interpretation
of these kinematic properties in \S\ref{sec:results.globalproperties}.

\subsection{Extinction}
\label{ssec:extinction}
The extended gas of powerful HzRGs is often very dusty, causing
several magnitudes of extinction in the rest-frame V band
\citep{Nesvadba2008}. We measure the H$\alpha$/H$\beta$ decrement to
estimate the extinction in the warm ionized gas of our sources,
assuming an intrinsic Balmer decrement of H$\alpha$/H$\beta$~=~2.9 and
adopting the galactic extinction law of \citet{Cardelli1989}. We
detect H$\alpha$ in all seven galaxies where it falls into the atmospheric
windows. For NVSS~J004000$-$303333 and NVSS~J234235$-$384526, which are at
z $\gtrsim$ 3.5, we cannot observe H$\alpha$ from the ground. H$\beta$
is detected in eight  of the nine integrated spectra of the galaxies of our
sample, and six galaxies have both lines in common. For
NVSS~J201943$-$364542, we can only set an upper limit to the H$\beta$
flux and, consequently, we give lower limits to the extinction.  We
find typical extinctions between formally $A_{H\beta}=$~0~mag in four
galaxies, and up to $A_{H\beta}=$~1.9~mag, where $A_{H\beta} = A_V -
0.14$ mag. Results for individual sources are listed in
Tab.~\ref{table:ExtinctionElectronDensityMassEnergy}.

\subsection{Electron densities}
The ratio of the lines in the $[$SII$]$$\lambda\lambda$6716,6731
doublet is density-sensitive over large ranges in density from about
100~cm$^{-3}$ to $10^5$~cm$^{-3}$ and can be used to estimate the
electron density in the emission-line gas
\citep{Osterbrock1989}. These lines are well detected in two galaxies,
 NVSS~J012932$-$385433, where they are broad and blended, and in
NVSS~J210626$-$314003 (see Fig.~\ref{fig:spec1}
and~\ref{fig:spec2}). We find $n_e = 750$~cm$^{-3}$ for
NVSS~J012932$-$385433 and $n_e = 500$~cm$^{-3}$ for
NVSS~J210626$-$314003, assuming a temperature $T = 10^4$~K. 

Electron densities of 500-700 cm$^{-3}$ are higher by factors of a few 
than those in low-redshift AGNs with powerful radio sources and
electron densities of a few 10 to about 100 cm$^{-3}$
\citep[e.g.,][]{stockton76}. This mirrors the higher electron
densities of a few~100~cm$^{-3}$ in the interstellar gas of star-forming
galaxies at z$\sim$2 \citep[][]{letiran11} compared to starburst
galaxies in the nearby Universe, and it also explains the higher surface
brightnesses of extended gas in our galaxies compared to low-z AGN
hosts. Similar electron densities of a few 100~cm$^{-3}$ have previously
been found in other HzRGs \citep[e.g.,][Nesvadba et al., 2015, in
  prep.]{Nesvadba2006, humphrey08, Nesvadba2008}, but we caution
nonetheless that the value we adopt here is uncertain by factors of a
few. We adopt a fiducial value of $n_e = 500$~cm$^{-3}$ for the other
galaxies in our sample, which corresponds to the average of HzRGs with
appropriate measurements.

\subsection{Ionized gas masses}
\label{ssec:gasmass}
Estimating the mass of warm ionized gas in high-redshift galaxies is
straightforward, and can be done by measuring the flux of the bright
Balmer lines and the electron density in the gas. The basic principle
of the measurement is to count the number of recombining photons at a
given electron density. Assuming case B recombination, we can estimate
the ionized gas mass following \cite{Osterbrock1989} by setting

\begin{equation}
M_{ion} = 3.24 \times 10^{8}\frac{L_{H\alpha}}{10^{43} {\rm erg s}^{-1}} \frac{10^2 {{\rm cm}^{-3}}}{{\rm n_{e}}} M_{\odot}
,\end{equation}
where $L_{H\alpha}$ is the H$\alpha$ luminosity
corrected for extinction and $n_e$ is the electron density.

We find masses of ionized gas in our sample in the range $1 - 10
\times 10^8$~M$_\odot$ when using extinction-corrected luminosities,
and in the range $0.5 - 5 \times 10^8$~M$_\odot$ when using the observed
luminosities of H$\alpha$ without taking extinction into account.
This is generally less than in previous studies of more powerful radio
galaxies \citep[e.g.,][Nesvadba et al., 2015, in
  prep.]{Nesvadba2006,Nesvadba2008}, which have masses of warm
ionized gas between $10^9$ and a few $\times 10^{10}$~M$_\odot$.

\section{Properties of AGNs and black holes }
\label{ss:agnproperties}

\subsection{Centimeter radio continuum}
\label{sec:radiocontinuum}

Our sources cover a range of radio sizes and morphologies, ranging
from single compact sources of the size of the ATCA beam (typically
1\arcsec\ $-$3\arcsec\ at the highest observed frequency of 9~GHz) and
single compact sources that potentially have faint extended
structures to doubles with sizes of up to 25\arcsec. Only in one case
do we potentially detect the radio core along with the lobes. This
spatial resolution is lower than  can be achieved with the JVLA at
the highest resolution, but most of our targets are too far south to
be observed with the JVLA with a good, symmetric beam.  Nonetheless,
the resolution of these data is sufficient to distinguish between
radio sources that have and have not yet broken out of the ISM of
the host galaxy (i.e., sources with sizes below or above approximately
10~kpc), which is the most important distinction for our purposes.

We use our observed radio fluxes, previous ATCA results from
\citet{Broderick2007} and \citet{Bryant2009b}, and the NVSS results
from \citet{Brookes2008CENSORS} to constrain the radio power at
different frequencies, the radio spectral index, and the kinetic
energy of the radio lobes. In Tab.~\ref{tab:atcaresults} we list the
radio power of each source at a rest-frame frequency of 1.4~GHz, which
is frequently given in the literature for high-redshift quasars
\citep[e.g.,][]{Harrison2012} and at 500~MHz in the rest-frame, which
is the reference frequency of the sample of powerful radio galaxies at
z$\sim$2 in the compilation of \citet{miley08}.

With a rest-frame radio power at 1.4~GHz between $3\times 10^{26}$ W
Hz$^{-1}$ and $8\times 10^{27}$ W Hz$^{-1}$ ($8\times 10^{26}$ W
Hz$^{-1}$ to $3\times 10^{28}$ W Hz$^{-1}$ at 500~MHz,
Tab.~\ref{tab:atcaresults}), our sources are intermediate between
typical dusty starburst galaxies and the most powerful radio galaxies
at similar redshifts. To calculate radio spectral indices, we used our
own measurements at 5.5~GHz and 9.0~GHz, 1.4~GHz observations from the
NVSS catalog, and for the MRCR-SUMSS sample the \citet{Broderick2007}
results at 0.408~GHz, 0.843~GHz, and 2.368~GHz. Spectral indices are
estimated from best fits to the power law of the radio spectral energy
distribution, giving values between $\alpha=-0.8$ and $-1.4$, without
clear evidence of spectral breaks. Table~\ref{tab:atcaresults} lists
the results for each individual source. For the MRCR-SUMSS sample, our
results agree with those of \citet{Broderick2007}. For the CENSORS
sample, no previous measurements of the radio spectral index were
available.

We use the measured radio fluxes and spectral indices to extrapolate
the radio power down to the rest-frame frequencies for which empirical
calibrations of the kinetic energy of the radio source have been
derived. The observed jet power is only a small fraction ($\le 1$\%)
of the mechanical power of the radio jet. Specifically, we use the
relationship of \citet{Willott1999} which is based on the 151~MHz flux
in the rest frame (and given in units of $10^{28}$ W Hz$^{-1}$
sr$^{-1}$), $L_{151,28}$, and set $L_{\rm jet} = 3 \times 10^{45}
f^{3/2} L_{151, 28}^{6/7} {\rm erg} \ {\rm s}^{-1}$, where $f$
represents the astrophysical uncertainties and is typically between 1
and 20. Here, we use $f = 10$ \citep[see also][]{cattaneo09}. 

We also use the calibration of \citet{Cavagnolo2010} which measures the
work needed to inflate the X-ray cavities found in low-redshift galaxy
clusters as a proxy to the mechanical power of the radio jet, $\log
P_{cav}= 0.75(\pm 0.14) \log(P_{1.4} + 1.91(\pm 0.18))$, where $P_{cav}$
is the kinetic power of the X-ray cavity given in units of $10^{42}$
erg s$^{-1}$, and $P_{1.4}$ the radio power at 1.4~GHz in the rest
frame. Both approaches give broadly similar results with differences
of typically about 0.1~dex for the MRCR-SUMSS sources, and differences
of 0.3~dex for the CENSORS sources, which have steeper spectral
indices. We list all results in Tab.~\ref{tab:atcaresults}.

\subsection{Bolometric AGN emission}

For a first estimate we consider the [OIII]$\lambda$5007 luminosity
(Tab.~\ref{table:ExtinctionElectronDensityMassEnergy}) as a signature
of the nuclear activity in our galaxies. Even if they neglect
extinction, our measured [OIII]$\lambda$5007 line fluxes indicate
luminosities in the range of 0.1$-$few $\times 10^{44}$~erg~s$^{-1}$,
in the range of powerful quasars and only somewhat fainter than the
[OIII]$\lambda$5007 luminosities of the most powerful HzRGs. As
  discussed in more detail in Nesvadba et al. (2015, in prep.), the
  line ratios in our galaxies are also consistent with being
  photoionized by their powerful AGNs. Correcting the fluxes for
extinction, which is relatively low in our targets (see
\S\ref{ssec:extinction}), increases these values by factors of a
few. Using the relationship of \citet{Heckman2004}, this would
correspond to bolometric luminosities of the AGNs on the order of  ${\cal
  L}_{bol}$ = 3500$\times {\cal L}$([OIII]) or a few $10^{46-47}$ erg
s$^{-1}$. Although this estimate is known to overestimate the
intrinsic bolometric luminosities of radio-loud quasars by factors of
up to a few, this does not change the result at an
order-of-magnitude level as we are stating here.

\subsection{Broad-line components and properties of black holes}
\label{ssec:BLR}

Constraining the AGN properties of high-z radio galaxies is
notoriously difficult since it is the very nature of radio galaxies
that the direct line of sight into the nucleus is obscured.
However, in a few fortuitous cases \citep[e.g.,][]{Nesvadba2011a},
broad H$\alpha$\ lines have been observed that are likely to trace
the gas kinematics within a few light-days from the supermassive
black hole \citep[][]{kaspi00,peterson04}. Such lines can be used to
constrain the mass of the black hole and its accretion rate
\citep[e.g.,][]{GreeneHo2005}.

As described in \S\ref{s:PresentationOfOurAnalysisTools}, we fitted
the spectra of our sources with single Gaussian components. Within the
uncertainties of our data this yields acceptable fits to the
integrated spectra of most targets and for most emission
lines. However, three of our targets, NVSS~J002431$-$303330,
NVSS~J012932$-$385433, and NVSS~J201943$-$364542, require additional
H$\alpha$ components. These components have FWHM
$\ge$ 3000~km s$^{-1}$, significantly broader than the systemic line
emission. Moreover, NVSS~J234235$-$384526 clearly shows a second
[OIII]$\lambda$5007 component. Fainter, marginally detected broad
[OIII]$\lambda$5007 components may also be present in
NVSS~J002431$-$303330 and NVSS~J004000$-$303333.

The origin of broad (FWHM $\gg$ 1000~km s$^{-1}$) H$\alpha$ line
emission at high redshift has either been attributed to winds driven
by starbursts \citep[e.g.,][]{leTiran2011b, Shapiro2009,Nesvadba2007b}
or active galactic nuclei on galaxy-wide scales
\citep[e.g.,][]{Alexander2010, Nesvadba2011a, Harrison2012}, or
alternatively, to gas motions in the deep gravitational potential
wells very near the supermassive black hole
\citep[e.g.,][]{Alexander2008, Coppin2008, Nesvadba2011a}. For
galaxies like NVSS~J234235$-$384526, and perhaps NVSS~J002431$-$303330 and
NVSS~J004000$-$303333, which only have  relatively broad [OIII]
components, it is clear that the broad-line emission probes gas in the
narrow-line region or outside, at larger galactocentric radii. This is
similar to high-z quasars \citep[e.g.,][]{Harrison2012,
  Nesvadba2011b}, since forbidden lines are collisionally de-excited
at the high electron densities of the broad-line region
\citep[e.g.,][]{Sulentic2000}.  The non-detection of these components
in the Balmer lines make it unlikely that these winds encompass large
gas masses. High electron densities and ionization parameters can
boost the luminosity of the [OIII]$\lambda\lambda$4959,5007 lines
without implying large gas masses (e.g., Ferland 1993).

For galaxies where we only observe broad components in the H$\alpha$
and [NII]$\lambda\lambda$6548,6583 complex, with widths that make it
difficult to uniquely associate the broad-line emission with either
line, the situation is less clear. Line widths of $\ge 3000$~km
s$^{-1}$ have been considered as evidence of BLRs in submillimeter
galaxies and quasars at z~$\sim$~2 \citep{Alexander2008, Coppin2008};
however, extended emission lines with FWHM $\sim$ 3500~km s$^{-1}$
have been observed in very powerful radio galaxies at similar
redshifts \citep[e.g.,][]{Nesvadba2006}, and given the generally large
line widths in our sources, it is clear that the ISM is experiencing a
phase of strong kinetic energy injection.

Among our sources, NVSS~J201943$-$364542 clearly stands out in terms
of line width and line luminosity. It is the only galaxy with a line
as broad as (FWHM = 8250~km s$^{-1}$) and as luminous as  (${\cal
  L}=1.1\times 10^{44}$~erg~s$^{-1}$)  the H$\alpha$ emission from
nuclear broad-line regions in powerful radio galaxies at similar
redshifts \citep{Nesvadba2011a}, and exceeds the ``typical'' bona fide
AGN-driven wind by nearly an order of magnitude in line width. We do
not find a broad component in [OIII] that would correspond to the one
seen in H$\alpha$.

Following \citet{GreeneHo2005} we can use the width and the luminosity
of the H$\alpha$ line to estimate the mass and accretion rate of the
supermassive black hole, M$_{\mathrm{BH}} = 2.1^{+2.2}_{-1.1} \times
10^9$ M$_{\odot}$ and $L_{bol,AGN} = 1.4\times10^{12} {\cal
  L}_{\odot}$ and $\mathcal{L}_{\mathrm{bol}} = 5.3 \times 10^{45}$
erg~s$^{-1}$. The Eddington luminosity of a M$_{\mathrm{BH}} =
2.1^{+2.2}_{-1.1} \times 10^9$ M$_{\odot}$ black hole is
$\mathcal{L}_{\mathrm{Edd}} = 2.8 \times 10^{47}$ erg~s$^{-1}$,
corresponding to an Eddington ratio of 2\%. These values are well
within the range found for more powerful radio galaxies at z~$\ge$~2
\citep{Nesvadba2011a}, including the remarkably low black hole
 Eddington ratio compared to many bright high-redshift quasars
\citep[see][for a detailed discussion]{Nesvadba2011a}. We caution,
however, that we did not take into account that the extinction might be
greater than that of optically selected quasars. Our non-detection of
H$\beta$ implies A(H$\beta$)~$\ge$~1.7, which would be less than the
$A_V=3.5$ mag used by \citet{GreeneHo2005}, if taken at face value. If
the unified model applies for these galaxies \citep[][]{antonucci93,
  drouart12}, then differences in inclination are likely the largest
uncertainty of about a factor of 2, with little impact on our
results. \citet{drouart14} find higher Eddington ratios from
Herschel/SPIRE observations of five of the \citet{Nesvadba2011a} sources
when adopting a bolometric correction factor of 6 between the
far-infrared and bolometric luminosity of these galaxies. The measured
far-infrared luminosities are lower by  factors of up to 5  than the ${\cal
  L}_{bol}$ derived from the H$\alpha$ line luminosity.

The more moderate line widths (FWHM $\ge$ 3000~km s$^{-1}$) and line
luminosities of NVSS~J002431$-$303330 and NVSS~J012932$-$385433 make
interpreting the nature of the broad-line components in these galaxies
more difficult, and it is not possible with the present data alone to
clearly distinguish between the wind and the black hole
hypothesis. Under the assumption that these lines probe the AGN
broad-line region, we find (using  the same approach as above and using
the measurements listed in Tabs.~\ref{tab:spec1}~and~\ref{tab:spec2})
$L_{bol,AGN}^{0129}= 6.4\times 10^{44}$~erg~s$^{-1}$ and
$L_{bol,AGN}^{0024}= 8.5\times 10^{44}$~erg~s$^{-1}$ for
NVSS~J012932-385433 and NVSS~J002431$-$303330, respectively, and
black hole masses of M$_{\mathrm{BH}}^{0129} = 2.8 \times
10^8$~M$_{\odot}$ and M$_{\mathrm{BH}}^{0024} = 2.4 \times
10^8$~M$_{\odot}$, respectively. The Eddington ratios would be as low
as for NVSS~J201943$-$364542, about 2$-$3\%. We caution, however, that we
have no unique constraint to distinguish between AGN broad-line
emission, and gas interacting with the radio jet on larger scales in
these two sources. If confirmed to be tracing BLR gas, these two
galaxies would have supermassive black holes with masses closer to
submillimeter galaxies than more powerful radio galaxies at
z~$\sim$~2, although their accretion rates would be significantly
lower than in submillimeter galaxies \citep[which accrete near the
  Eddington limit;][]{Alexander2008}. However, their kinetic jet power
\citep[$5\times 10^{46}$ erg s$^{-1}$ for both sources, using
  the approach of][]{Cavagnolo2010}, would slightly exceed their Eddington
luminosities, $L_{Edd}=3.6$ and $3.1\times 10^{46}$ erg s$^{-1}$, and
their bolometric luminosities implied by H$\alpha$ would be two orders
of magnitude lower than those implied by their [OIII]
luminosities. Although super-Eddington accretion is not impossible, it
is very rarely observed, which is why we are doubtful that these are
 bona fide AGN broad-line regions.

\section{Additional line emitters and dynamical mass estimates of our HzRGs}
\label{s:additionalLineEmittersTracersOfEnvironmentDynamicalProbes}

Many high-redshift radio galaxies do not exist in solitude. A large
number of imaging and spectroscopic studies have demonstrated
conclusively that many massive radio galaxies at z~$\ge$~2 are
surrounded by galaxy overdensities at the same redshift, as well as
significant reservoirs of molecular, atomic, and ionized gas,
including diffuse intra-halo gas \citep[][]{leFevre1996, Venemans2007,
  Hayashi2012, Galametz2012, vanOjik1997, VillarMartin2003,
  DeBreuck2003a, Nesvadba2009, wylezalek13, collet14}.  With the small
field of view of only 8\arcsec$\times$8\arcsec, SINFONI can only
constrain the very nearby environment of HzRGs out to a few tens of kpc
(8\arcsec\ correspond to 64~kpc at z~$\sim$~2); however, this
small-scale environment is particularly interesting, e.g., to study
how accretion and merging may affect the evolutionary state of the
radio galaxy. Given the presence of extended gas clouds well outside
the radio galaxy itself, which we attribute to AGN-driven winds and
bubbles, it may not be immediately clear how to identify a secondary
line emitter within a few tens of kpc from the radio galaxy as another
galaxy in each individual case.

We have three such examples, NVSS~J004000$-$303333, NVSS~J201943$-$364542,
and NVSS~J144932$-$385657. For the last, we argue in
\S\ref{ssec:rotation} why we favor the interpretation of an AGN-driven
bubble. For NVSS~J004000$-$303333 and NVSS~J201943$-$364542 the situation
is more difficult.  In either case, the line emission cannot be
geometrically associated with the radio jet, which strongly disfavors
a direct physical connection between jet and gas. The redshifts in
both cases are significantly offset from those in the radio galaxy
itself. In NVSS~J201943$-$364542, the line widths of this second
component are also very narrow, FWHM $=$ 320~km s$^{-1}$, and the
light is emitted from within about 1\arcsec\ (8~kpc at z~$=$~2.1), as
would be expected from a low-mass galaxy in the halo of
NVSS~J201943$-$364542. The high [OIII]/H$\beta$ ratios  observed in this
putative companion are consistent with the values observed in rest-frame
UV selected, fairly low-mass galaxies such as Lyman-break galaxies (LBGs)
and other blue, intensely star-forming galaxies at high redshifts. The
[OIII]$\lambda$5007 luminosity of our source, ${\cal L} = 5.3 \times
10^{42}$~erg~s$^{-1}$, does not stand out compared to  the LBGs
of \citet{Pettini2001}, for example,  which typically have ${\cal L} = {\rm a\ few}
\times 10^{42}$~erg~s$^{-1}$. 

In NVSS~J004000$-$303333, the line emission is closer to the HzRG,
within about 1\arcsec, and brighter, ${\cal L}$([OIII]) = $4\times
10^{43}$~erg~s$^{-1}$. This may indicate that the gas is at least
partially lit up by photons originating from the AGN in the radio
galaxy. The proximity to the radio galaxy  may also suggest that
this gas is already dominated by the gravitational potential of the
radio galaxy itself, either as part of a satellite galaxy that is
being accreted (e.g., Ivison et al. 2008, Nesvadba et al. 2007), or
perhaps because it is associated with an extended stellar halo forming
around the HzRG, as observed in other cases \citep[][see also
  \citealt{collet14}]{Hatch2009}

With both hypotheses, the projected velocity of these additional line
emitters would be dominated by gravitational motion, and can therefore
be used as an order-of-magnitude measure of the dynamical mass of the
central radio galaxy and its underlying dark matter halo. Assuming
that the system is approximately virialized, we set $M = v_c^2\ R /
G$, where the circular velocity, $v_c$, is $v_c = v_{{\rm obs}} /
\sin{i}$, and where $R$ is the projected radius, and $G$ the
gravitational constant.  With $v_{\rm obs} = 408$~km s$^{-1}$ for the
companion of NVSS~J201943$-$364542, $v_{\rm obs}=350$~km s$^{-1}$ for
NVSS~J004000$-$303333, and projected distances of 24~kpc and 8~kpc,
respectively, we find dynamical mass estimates of $9\times
10^{11}\ \sin^{-1}{i}$~M$_{\odot}$ for NVSS~J004000$-$303333 and of
$3\times 10^{11}\ \sin^{-1}{i}$~M$_{\odot}$ for NVSS~J201943$-$364542,
respectively. Both are in the range of stellar and dynamical masses
estimated previously for the most powerful HzRGs \citep{Seymour2007,
  deBreuck2010, Nesvadba2007b, VillarMartin2003}.

\section{Signatures of AGN feedback}
\label{sec:results.globalproperties}

\subsection{Comparison with other massive high-z galaxies} 
\label{ssec:buitrago}

Given the complexity of the gas kinematics of high-redshift galaxies
and incomplete sets of observational constraints, the astrophysical
mechanism that dominates the gas kinematics is not always easy to
identify {\it \emph{ab initio}}. It is therefore illustrative to
compare our sources with sets of other massive, contemporary galaxy
populations with imaging spectroscopy to highlight the peculiarities
of our galaxies in an empirical way. Specifically, we compare our
sample with the stellar-mass selected sample of \citet{buitrago13} of
ten galaxies with M$_{stellar}$$\ge 10^{11}$ M$_{\odot}$ at z$\sim$1.4
without obvious AGN signatures, and with the submillimeter selected
dusty starburst galaxies of \citet{alaghband12} and \citet{menendez13}
without very powerful AGN, and those of \citet{harrison12} and
\citet{alexander10} with powerful obscured quasars.

The \citet{buitrago13} and starburst-dominated submillimeter galaxies
(SMGs) generally have similar or even larger velocity gradients than
our sources, and these gradients are often very regular with a
monotonic rise from one side of the emission-line region to the
other. Total velocity offsets are between 200 and 600 km s$^{-1}$ in
the \citet{buitrago13} sample, and between 100 and 600 km s$^{-1}$ in
the SMGs of \citet{alaghband12} and \citet{menendez13}. In this
comparison we discard one galaxy of \citet{alaghband12} which shows
AGN characteristics in the optical, but not in the far-infrared.  The
sizes of emission-line regions in both samples are between
1\arcsec\ and 3\arcsec.

A significant difference between our HzRGs and the comparison samples
are, however, the highest FWHMs in our sources, between 400 km
s$^{-1}$ and 1500 km s$^{-1}$. FWHMs in the \citet{buitrago13} sample
are 100$-$300 km s$^{-1}$, and in the starburst-dominated SMGs they are
between 160 and 470 km s$^{-1}$. Even in the SMGs, where the gas kinematics
may be severely affected by ongoing galaxy mergers, the line
widths are significantly lower than in our galaxies. 

Larger FWHMs than in the mass-selected and starburst-dominated sources
are also found by \citet[][see also
  \citealt{Alexander2010}]{Harrison2012}, who study the ionized gas
kinematics in eight sub-mm selected starburst/AGN composites at
z~$\sim$~1.4$-$3.4 with imaging spectroscopy of
[OIII]$\lambda$5007. Their galaxies are detected at 1.4~GHz with
luminosities of $10^{24-25}$~W~Hz$^{-1}$, too low to disentangle the
contribution of star formation and radio source. Bolometric
luminosities of their obscured AGN are a few $10^{46-47}$ erg s$^{-1}$,
similar to powerful radio galaxies.  They find FWHM=200$-$1500 km
s$^{-1}$, not very different from our targets, and total velocity
offsets of 100$-$800 km s$^{-1}$.

However, their galaxies have integrated [OIII] line profiles that are
very different from those in our sample. Where signal-to-noise ratios
permit, their lines show a narrow component superimposed onto a broad
line, often with widths well in excess of FWHM $\sim$ 1000~km
s$^{-1}$. Similar [OIII] profiles were reported by
\citet{Nesvadba2011b} and \citet{CanoDiaz2012} for other obscured
quasars at similar redshifts. In turn, our HzRGs have one, generally
broad, line component with FWHM$=$500$-$1000 km s$^{-1}$. This
suggests that significant parts of the gas in high-z quasar hosts are
not strongly perturbed by the AGN. \citet{Nesvadba2011b} find that the
broadest line widths are found in an unresolved region near the AGN,
where [OIII] luminosities are likely boosted by high electron
densities.  Consistent with this finding, the quasars of
\citet{Harrison2012} exhibit velocity curves that are fairly regular
and flattened at low surface brightness, consistent with the turnover
expected for rotation curves. Although our sources do have extended
lower surface-brightness line emission with more quiescent kinematics,
their contribution to the integrated line profile is smaller, and this
gas is outside the region where a connection with the radio source is
obvious \citep[][]{villar03}. In quasars and Seyfert galaxies at low
redshift, \citet{husemann13} and \citet{mullersanchez11} also find
that the gas is only perturbed near the radio source, in broad
agreement with our findings \citep[but see][]{liu13}.

\subsection{Outflows or disk rotation?} 
\label{ssec:rotation}
The large velocity offsets of up to 2500 km s$^{-1}$ in the most
powerful HzRGs make it relatively easy to conclude that the gas is
driven by powerful, non-gravitational processes
\citep[][]{Nesvadba2006,Nesvadba2008}. In the sources we are studying
here, the situation is not so clear. Although we do find total
velocity offsets of a few 100 km s$^{-1}$ in the six galaxies with
resolved gas kinematics, in four cases they are relatively small, $\Delta
v=$ 300-450 km s$^{-1}$, and, as we just discussed in
\S\ref{ssec:buitrago}, within the range found in massive early-type
galaxies at high redshift without powerful AGNs. If we assume that this
gas is in a rotating disk and derive a mass estimate from the virial
theorem
(\S\ref{s:additionalLineEmittersTracersOfEnvironmentDynamicalProbes}),
this corresponds to masses of $0.5-1.2\times 10^{11}$ M$_{\odot}$ for
typical velocities $v/sin\ i =$150$-$225 km s$^{-1}$, radii R$=$5 kpc
(corresponding to a typical observed disk radius of 0.6\arcsec), and
an average inclination of 45$^\circ$ \citep[][]{drouart12}.  This is
similar to the expected mass range of a few $10^{11}$ M$_{\odot}$ of our
targets that we derived previously
(\S\ref{s:additionalLineEmittersTracersOfEnvironmentDynamicalProbes}),
and within the stellar mass range of powerful HzRGs of a few $10^{11}$
M$_{\odot}$ \citep[][]{seymour07, debreuck10}. Moreover, the existence
of kpc-scale rotating disks in a few low-redshift radio galaxies
\citep[e.g.,][]{emonts05, Nesvadba2011c} suggests that disk rotation
is very possible in galaxies with radio sources not much weaker than
we find here.

It is  interesting, however,  that our sources have at the same time
smaller velocity gradients and larger nebulosities than in the
mass-selected sample of \citet{buitrago13}. If the velocity
gradients we observe are indeed due to rotation, then this would imply
that our radio galaxies have  shallower mass profiles, which might
point toward a different mass assembly history in the two  populations. A
similar situation is  found, however,  in the radio-loud,
intermediate-redshift quasars of \citet{fu09}, which also have gas
that is relatively extended compared to the size of the stellar
component of their galaxies, and probes velocity ranges that are
smaller than those expected from stellar velocity dispersions.  They
conclude that the gas kinematics are unlikely to be directly related
to rotational motion. At high redshift we cannot probe a similarly
clear tracer of galaxy kinematics such as stellar absorption line
widths, and therefore cannot address this question directly.

Additional arguments point toward the outflow hypothesis.  The
extended emission-line regions in all but one source \citep[discussed
  further in][]{collet14} are aligned with the radio jet axis within
20$^\circ-30^\circ$ as would be expected if the jet were inflating an
overpressurized bubble of hot gas. Moreover, hydrodynamic models of
jet-driven winds do predict that jets with kinetic power of
$10^{46-47}$ erg s$^{-1}$ should accelerate gas to about 500 km
s$^{-1}$, which is consistent with our findings, in particular if we
take seeing (\S\ref{ssec:kinematics}) and projection effects into
account (see also \S\ref{ssec:wagner}).

\begin{figure}
\centering
\includegraphics[width=0.8\textwidth]{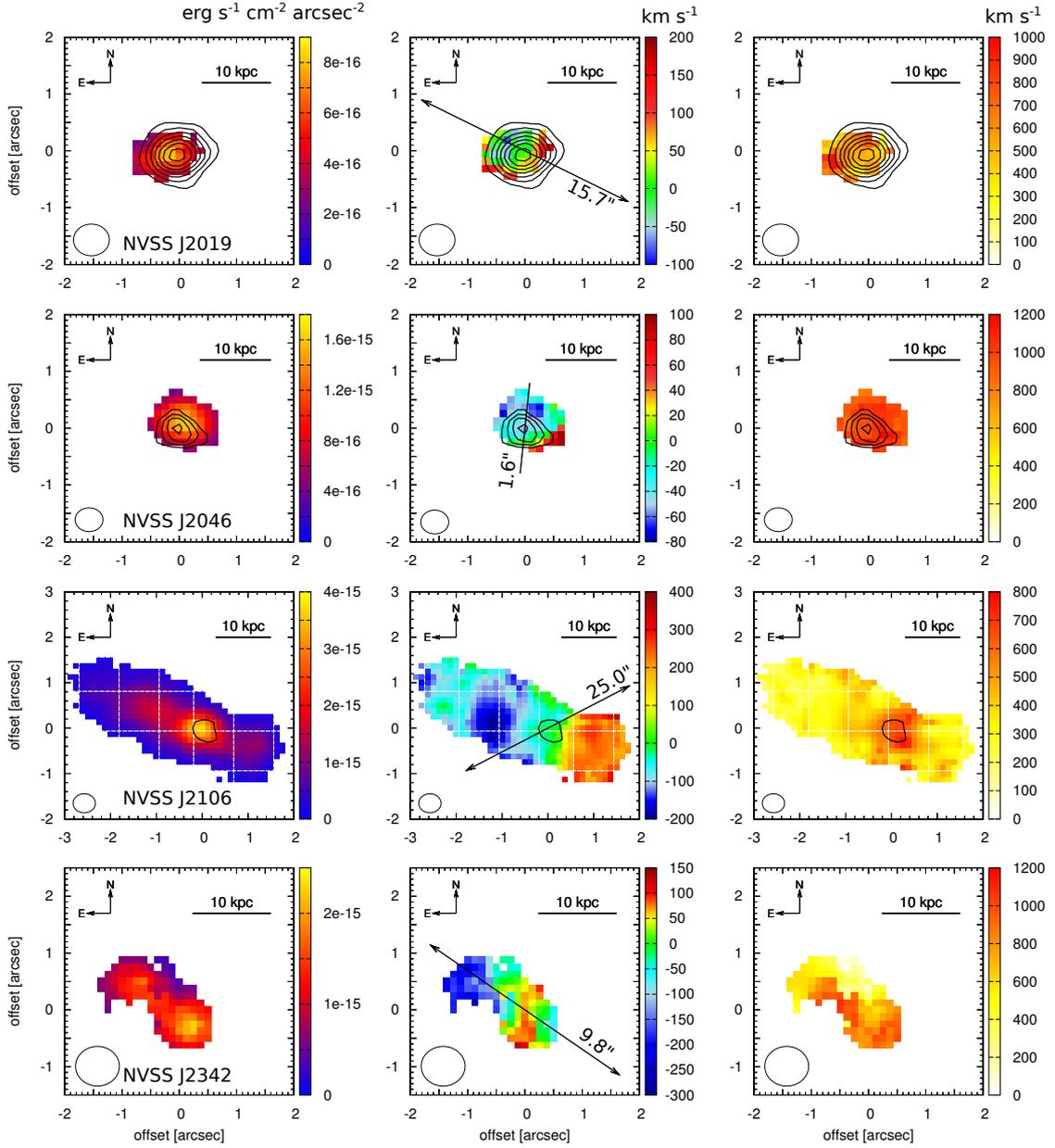}
\caption{Maps of [OIII] surface brightness ({\it left}), velocity
  ({\it center}), and FWHM line widths ({\it right}).  All maps are
  4\arcsec$\times$4\arcsec\ wide, except for NVSS~J144932$-$385657 and
  NVSS~J210626$-$314003, where they are
  5\arcsec$\times$5\arcsec\ wide. The circle in the bottom left
  represents the FWHM size of the seeing disk. Contours mark the
  continuum where detected. Continuum levels begin at 3$\sigma$ and
  then increase in steps of 1$\sigma$.  The solid black line in the
  velocity maps indicates the axis of the radio emission from our ATCA
  data, or from \citet{Broderick2007} if the source is compact. Numbers
  give the largest angular size in arcsec.
\label{fig:maps2}}
\end{figure}

\begin{figure}
\begin{center}
\includegraphics[width=0.8\textwidth]{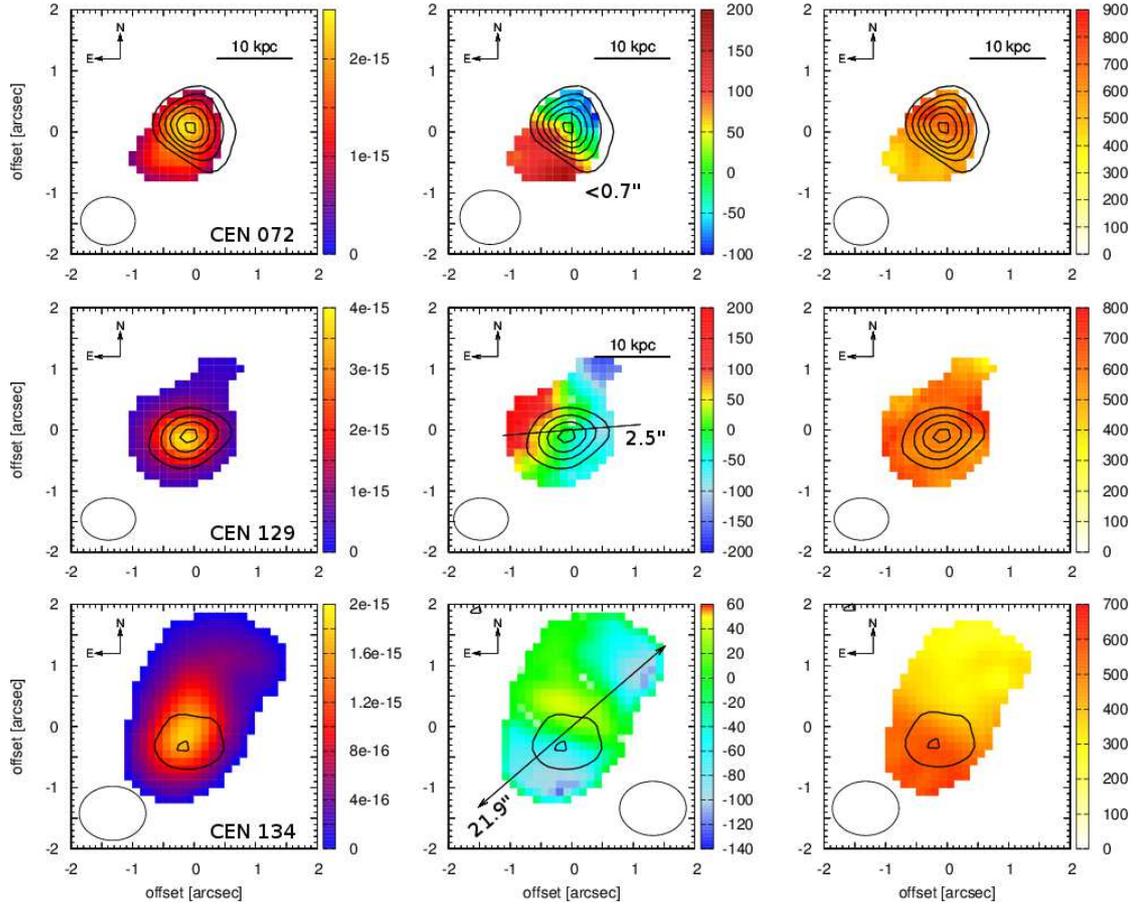}
\caption{Maps of [OIII] surface brightness ({\it left}), velocity
  ({\it middle}), and FWHM line width ({\it right}) of the three sources
  from the CENSORS sample. All maps are 4\arcsec$\times$4\arcsec\ on
  each side. The ellipse in the bottom corner shows the FWHM size of
  the seeing disk. Contours mark the stellar continuum emission,
  detected in the same datacube, beginning at the 3$\sigma$ level and
  increasing by steps of 3$\sigma$.\label{fig:maps3}}
\end{center}

\end{figure}

\begin{figure}
\centering
\includegraphics[width=0.8\textwidth]{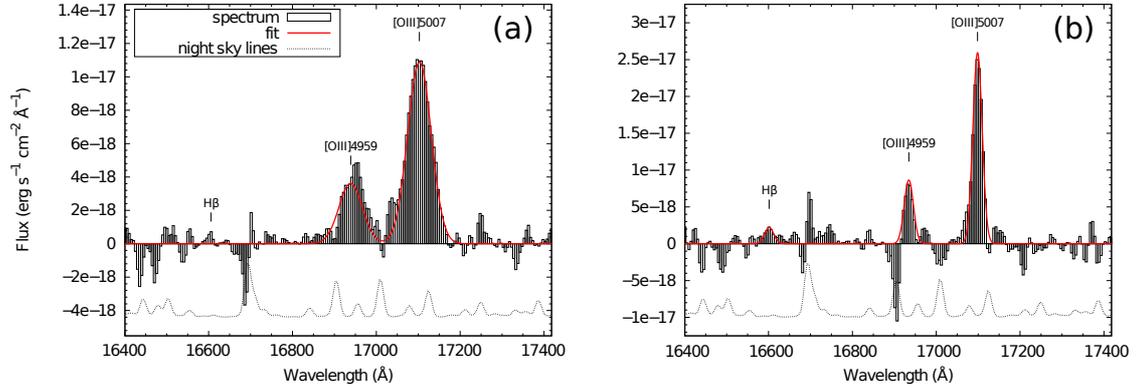}
\caption{Spectra extracted from the subregions of NVSS~J002431$-$303330 shown in
  Fig.~\ref{fig:spec1}. (a): Spectrum from the continuum
  region, with broad [OIII] lines (FWHM~$\sim$~1150~km s$^{-1}$). (b):
  Spectrum of the quiescent gas, with much more narrow lines
  (FWHM~$\sim$~450~km s$^{-1}$).\label{fig:J0024_OIIISpectra}}

\end{figure}

\begin{figure}
\centering
\includegraphics[width=0.8\textwidth]{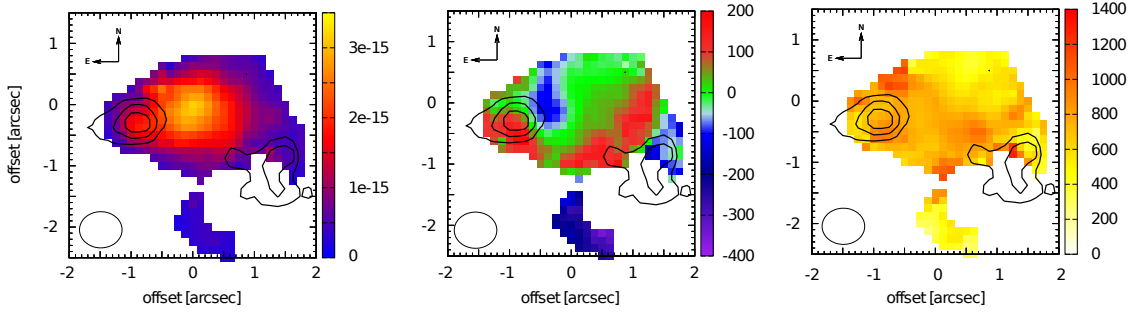}
\caption{Maps of NVSS~J004000$-$303333 at SNR~$\ge$~3.0. A second emission-line region clearly appears $\sim$~2\arcsec\ to the south of
  NVSS~J004000$-$303333, blueshifted by $350 \pm 90$~km s$^{-1}$ relative to
  NVSS~J004000$-$303333.\label{fig:J0040_SN3}}
\end{figure}

\begin{figure}
\centering
\includegraphics[width=0.8\textwidth]{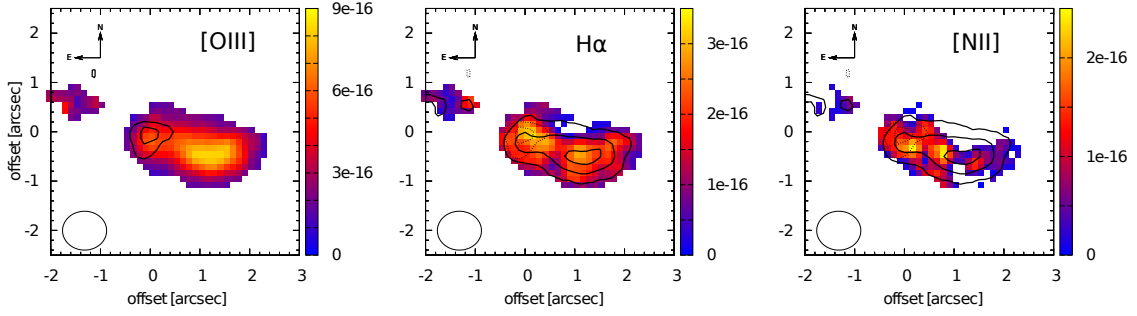}
\caption{Surface brightness maps of NVSS~J144932$-$385657 at
  SNR~$\ge$~3.0 for the [OIII]$\lambda5007$, H$\alpha$, and
  [NII]$\lambda6583$ emission lines. The contours in the left panel
  show the continuum and in the other panels the [OIII] emission-line
  morphology.\label{fig:J1449}}
\end{figure}

Most galaxies have velocity maps with significant irregularities even
at our relatively low spatial resolution of several kpc. Inspection of
Figs.~\ref{fig:maps1} to~\ref{fig:maps3} shows that all but perhaps
three sources, NVSS~J012932$-$385433, NVSS~J204601$-$335656, and
CEN~072, have significant residuals from simple, smooth, monotonic
velocity gradients.  This implies significant non-circular motion with
sudden velocity jumps of up to several 100 km s$^{-1}$, over scales of
a few kpc. A good example is NVSS~J002431$-$303330, where we see two
areas that are redshifted relative to the surrounding gas by about 200
km s$^{-1}$, or CEN~134, which shows two large regions toward the
northwest and southeast, which are blueshifted by up to about
150$-$200 km s$^{-1}$ relative to a central ridge of more redshifted
gas. It is clear that such kinematics cannot arise from simple disk
rotation.

We also find very extended gas in some cases, in particular in
NVSS~J144932$-$385657, where the line emission extends over 30~kpc,
clearly larger than an individual galaxy. The velocity offset between
the two bubbles is 800 km s$^{-1}$, and the gas has FWHM line widths
of up to about 800 km s$^{-1}$. The K-band image of
\citet{Bryant2009a} shows fuzzy, clumpy structures at the sensitivity
limit of the data near the eastern emission-line region, but no
single, clearly detected continuum source consistent with a galaxy. At
the depth of the image, we would have likely missed galaxies with
masses significantly lower than that of the radio galaxy; however, it
is not clear in this case how the lopsided morphology of the southern
bubble relative to the continuum of the radio galaxy, nor the fairly
high gas surface brightness in this putative small galaxy could be
explained. We therefore consider this source as an example of an
AGN-driven wind, similar to other HzRGs. The eastern bubble is much
smaller than the western bubble; however, similar asymmetries are
found in more powerful radio galaxies at the same redshift (Nesvadba
et al. 2015 in prep.). Moreover, the eastern blueshifted bubble is
near the edge of our data cube and may extend farther outside the
field of view.

In CEN~134, NVSS~234235$-$384526, NVSS~J144932$-$385657, and
NVSS~J004000$-$303333, the gas is elongated along the radio jet axis,
and in NVSS~J002431$-$303330, NVSS~J004000$-$303333,
NVSS~J030431$-$315308, NVSS~J144932$-$385657, NVSS~J234235$-$384526,
and CEN~134 the largest velocity offsets are also found near that
direction (Figs.~\ref{fig:maps1} to~\ref{fig:maps3}). This is
reminiscent of the ``jet-cloud interactions'' found in radio galaxies
near and far \citep[e.g.,][]{tadhunter91} and even in AGNs with low
radio power \citep[][]{fu09,mullersanchez11,husemann13} in the nearby
Universe.

Given the small velocity gradients and relatively small gas masses,
only a very small fraction of the kinetic jet power would be needed to
accelerate the gas to the observed velocities, even if all of the gas
were participating in outflows. We approximate the total observed bulk
kinetic energy of the gas simply by summing over the bulk kinetic
energy in each spatial pixel

\begin{equation}
\mathrm{E_{bulk}} = \frac{1}{2} \times \mathrm{M_{ion}^{corr}} \: \sum_{i \ \in \ \mathrm{EELR}} \frac{\Sigma(i)}{\Sigma_{\mathrm{tot}}} \times v(i)^2
,\end{equation}
where $M_{ion}$ is the mass of warm ionized gas estimated in
\S\ref{ssec:gasmass} corrected for extinction, and $v$ the velocity
offset in each pixel from the central velocity, which we consider an
acceptable approximation of the systemic velocity of the galaxy. The parameters 
$\Sigma(i)$ and $\Sigma(tot)$ are the gas surface brightness in each
spatial pixel and the total line surface brightness,
respectively. Measuring the energy in each spatial pixel allows us to
take the irregular gas kinematics into account. To probe the disk
kinematics out to faint surface brightness, we use the gas kinematics
as measured from [OIII], and scale by the H$\alpha$/[OIII] ratio in
the integrated spectrum. In galaxies where both lines are bright
enough to be probed individually, their surface brightness and
kinematics are sufficiently similar to justify this approach
\citep[see also][]{nesvadba08}.

With the velocities and warm ionized gas masses in
Tab.~\ref{table:Energy}, we find values between
$E_{bulk,min}=0.2\times 10^{56}$ erg and
$E_{bulk,max}=1\times 10^{57}$ erg in bulk kinetic energy in these
galaxies. This corresponds to a small fraction of the mechanical
energy carried by the radio jet,  a few percentage points or less
(Tab.~\ref{table:Energy}), if we assume typical jet lifetimes on the order of
$10^{6-7}$ yrs. This age range is expected from the dynamical time of
our radio jets, assuming a jet advance speed of $0.1 \ c$
\citep{Wagner2012}. It is consistent with the general range of a few
$10^6$ yrs found by \citet{blundell99} from spectral aging
considerations of powerful HzRGs \citep[see also][]{kaiser97}.

Thus, energetically, it would  be possible for jets with the
observed kinetic power to accelerate the gas to the observed
velocities. In particular, this is  the case for J144932$-$385657, where
the bulk kinetic energy amounts to 4\% of the kinetic energy carried
by the radio jet.

\subsection{Random motion and kinetic energy}

The large line widths are perhaps the most outstanding kinematic
property of the gas in our galaxies. They are similar to those found in
the most powerful radio galaxies at z$\sim$2 \citep[][]{nesvadba08},
and greater by factors of 2$-$3 than in other types of massive high-z
galaxies. Each source shows a range of line widths, and we
  deliberately compare them with the broadest widths near the nucleus
  because we want to quantify the impact of the AGN on the
  surrounding gas, and this is the gas that we expect to be most
  affected by the radio jet. Comparing the amplitudes of the
velocity gradients, $\Delta v/2$, with Gaussian line widths
$\sigma=FWHM/2.355$, we generally find ratios of $v/\sigma$
$=$0.3$-$0.8 (NVSS~J144932$-$385657 has a value of 2.3), compared to
$v/\sigma =$1$-$3.5 in the sample of \citet{buitrago13}. Rotationally
supported disk galaxies in the nearby universe typically have
$v/\sigma \sim$10. Individual values of our sources are listed in
Tab.~\ref{table:ExtinctionElectronDensityMassEnergy}.

At z$\sim$2 we cannot infer directly whether these line widths reflect
spatially unresolved velocity offsets on smaller scales, strong
turbulent motion, or a combination of both. In all cases, this suggests the
presence of an additional source of kinetic energy in our galaxies
that is stirring the gas up, and which is absent in the general
population of very massive high-z galaxies.

Finding v/$\sigma\lesssim$1 also implies  that the gas, even if it is
in a rotating disk, cannot be in a stable configuration. Except for
implausibly high inclination angles, gas in the line wings is at
velocities above the local escape velocity from the disk, and is
therefore not gravitationally bound. Nevertheless, most of the gas may
be bound to the galaxy itself, since the escape velocity of galaxies
of a few $10^{11}$ M$_{\odot}$ is about 700 km s$^{-1}$
\citep[][]{nesvadba06}. This would imply that most of the gas that is
being lifted off the disk will slow down as it rises to larger
galactic radii, and ultimately rain back toward the center of the
radio galaxy \citep[e.g.,][]{Alatalo2011}. This is particularly the case
if turbulence is indeed the cause of the line broadening, since
turbulent dissipation times are very short \citep[][]{maclow99}. 

Disks with low v/$\sigma$ values that are highly turbulent and not
gravitationally bound have been studied in a few individual cases at
low redshift \citep[][]{Alatalo2011,Nesvadba2011c}, and are
characterized by large line widths and complex line profiles as we
find here. These disks are very different from classical thin disks, for example, 
in  spiral galaxies. The complex line profiles and low volume
filling factors suggest the gas is generally filamentary and diffuse,
and cannot form gravitationally bound clouds and stars. Densities even
in the molecular gas traced by CO millimeter line emission are only
on the order of a  few 1000 cm$^{-3}$ \citep[][]{nesvadba10}, not very different
from those we find in the ionized gas of HzRGs \citep[][]{collet14,
  nesvadba08}. To understand the peculiar properties of these disks,
in particular the absence of clear signatures of ongoing and past
star formation, \citet{Nesvadba2011c} proposed that the dense gas in
these disks may have formed from the diffuse ISM through the pressure
enhancement in the cocoon inflated by the radio jet. Although this
scenario requires more observations at low redshift to be confirmed,
the broad properties of our targets may suggest that they may be
fundamentally similar to these gas-rich radio galaxies in the 
nearby Universe.

It is interesting to compare the kinetic energy in random motion with
that in bulk motion. To constrain the energy of random motion (which
we  loosely refer to as turbulent energy) we set 
\begin{equation}
\mathrm{E_{turb}} = \frac{3}{2} \times \mathrm{M_{ion}^{corr}} \sum_{i \ \in \ \mathrm{EELR}} \frac{\Sigma(i)}{\Sigma_{\mathrm{tot}}} \times \sigma(i)^2
\end{equation}
with velocity dispersion $\sigma$, i.e., FWHM/2.355. With the line
widths measured previously, we find values in the range
$E_{turb,min}=2.6\times 10^{57}$ erg and $E_{turb,max}=31\times 10^{57}$ erg in
turbulent kinetic energy. For typical jet ages of a few $10^{6-7}$ yrs,
this corresponds to energy injection rates of a few $10^{42}$ to
$10^{44}$ erg s$^{-1}$ (our precise numbers are for a fiducial $10^7$
yrs). The turbulent energy corresponds to a few tenths of a percent  to a small percentage of the
mechanical energy carried by the radio jet.

Under the assumption that bulk velocities are solely from radial
motion, ratios between bulk and turbulent kinetic energy are typically
between 0.2 and $<$10\%, only NVSS~J144932$-$385657 with very extended
gas has a ratio of 30\%. The small contribution of the bulk kinetic
energy to the total kinetic energy budget of the gas implies that the
uncertainties owing to the unknown split between rotational and
outflow motion do not have a large impact on the total kinetic energy
budget of the ionized gas. Finding that the largest ratio of bulk to
turbulent motion is in the galaxy with the most extended emission-line
regions, NVSS~J144932$-$385657 is interesting. More extended
emission-line regions would be consistent with an outflow where
dissipational losses through turbulence or random motion are less
important.

Summing the bulk and turbulent kinetic energy, we find that the AGN
deposits 3-10\% of the jet kinetic energy in the
gas. This is similar to the range found in very powerful HzRGs
\citep[][]{nesvadba08}, but a significant difference is that the
energy in random motion appears to be an order of magnitude greater
than that in ordered motion (on kpc scales), whereas both are roughly
equal in very powerful sources. 

Depending on the nature of the `turbulent' motion, this can have one
of two consequences. Either the line broadening is caused by strong
velocity changes over very small scales, in which case we may
underestimate the intrinsic velocity of the gas because of beam
smearing effects, and perhaps blending of high-velocity gas components
with more quiescent gas in an underlying disk. Low-redshift
equivalents include the compact radio galaxies of
\citet{holt08}, for example, or the broadening is indeed from turbulent motion, in
which case even fairly high-velocity gas may decelerate rapidly
(because turbulent dissipation times are likely short
\citep[][]{maclow99}) and rain back onto the disk.

It appears uncertain in either case whether significant parts of the
observed gas will escape from these galaxies in the current radio
activity cycle, unless our radio sources  undergo phases of
significantly higher power during the current activity cycle, and we
have merely observed them in an atypically low-power phase.

In either case, our findings are at odds with the simple scenario
whereby AGN feedback acts mainly by removing the gas
\citep[e.g.,][]{dimatteo05}. A cyclical model whereby gas  cools
down and accumulates at small radii near the supermassive black hole
before igniting another feedback episode, as has previously been
suggested for the central galaxies of massive galaxy clusters
\citep[][]{PizzolatoSoker2005}, may therefore be more appropriate
here. Statistical evidence for recurrent AGN activity has recently
also been discussed by \citet{hickox14}.
  
\subsection{Comparison with hydrodynamic jet models}
\label{ssec:wagner}
We will now compare our model with the hydrodynamic models of
\citet{Wagner2012} who quantify the energy transfer from jets into the
ambient gas in radio jet cocoons. They assume a clumpy, fractal
two-phase medium with a density distribution set by interstellar
turbulence like that found by \citet{padoan11}. The range in jet
power they cover with their 29 models is $10^{43-46}$ erg s$^{-1}$,
well matched to our sources, and Eddington ratios $\eta=10^{-4}$ to $1$.

They find that 30\%\ of the jet energy is transferred into the
gas, mainly through ram pressure transfer of partially thermalized
gas streams occurring in low-density channels between denser
clouds. This energy transfer corresponds approximately to the ratio
of gas kinetic to jet mechanical energy that we measure in our
data. The values found in the simulations are somewhat higher than
the values we find; however, our data can only be a lower limit
to the actual energy transfer since we do not observe all gas phases.
In particular the hot, X-ray emitting gas and the molecular gas are
missing. Likewise, uncertainties may arise from the details of the
density distribution of the gas, filling factors, etc., since the ISM
properties of high-redshift galaxies are not very well constrained
yet. In particular, cloud sizes -- a parameter that is not
observationally constrained at z$\sim$2 -- appears to play a major
role \citep[][]{Wagner2012}. Overall, given  these uncertainties, we
consider the correspondence between the model and data of an energy
transfer in the range of a few percentage points  very encouraging.

This correspondence is also illustrated in Fig.~\ref{fig:wagnerfig},
which was inspired by Fig.~11 in \citet{Wagner2012}. It shows the
expected gas velocities as a function of radio power for a range of
Eddington ratios. Our jets span the range of about $10^{46-47}$ erg
s$^{-1}$, which for Eddington ratios of 0.1 to 0.01 correspond to
velocities of about 250$-$500 km s$^{-1}$. This is somewhat higher
than the velocity offsets we observe, which might in part be
attributable to orientation effects and beam smearing. 

However, what we plot in Fig.~\ref{fig:wagnerfig} are not the velocity
offsets, but the Gaussian widths of the integrated spectral lines,
i.e., the overall luminosity-weighted range of velocities from random
gas motion.  This may further underline that a large part of the
energy injected by the radio jet in our galaxies does not result in an
ordered, large-scale outflow, but is either causing local small-scale
bulk motion or is ultimately being transformed into turbulent motion.
As \citet{Wagner2012} point out, estimating the long-term behavior of
the gas kinematics over kpc scales is not possible with the current
set of simulations, which only follow the evolution of the gas in the
first $10^3$ kyrs.

The hatched area in Fig.~\ref{fig:wagnerfig} indicates the range of
velocity dispersions measured by \citet{buitrago13} and illustrates
that for jets less powerful that those of our sources, it will be
difficult to identify the fingerprints of high-redshift AGNs with
observations of the kinematics of warm ionized gas in the presence of
other processes.

\subsubsection{Radiative quasar feedback?}

Another possible feedback mechanism that has recently been widely
discussed in the literature is radiative feedback from the bolometric
energy output of AGNs. It is attractive to explain the black hole
bulge scaling relationships via radiative processes because the
accretion rates implied by number counts of optically selected quasars
appear to be a good match to the local black hole demographics
\citep[e.g.,][]{yu02}. In spite of recent claims of AGN-driven bubbles
and winds in bright, low-redshift quasars \citep[e.g.,][]{liu13},
statistical evidence is still rather scarce. For example, so far it
has not been possible to find deviations in the star formation rates
of X-ray selected AGN hosts and galaxies without AGNs
\citep[e.g.,][]{mullaney12}.

Radiation pressure has received particular attention in the recent
literature.  Each time a photon scatters on a dust (or gas) particle
in the ISM it produces a small recoil in the ISM particle, with a net
momentum transfer for fluxes that are high enough, in particular when
the gas is optically thick, so that many interactions happen per
photon. \citet{murray05} derived analytic equations to approximate the
expected gas velocities as a function of quasar luminosity. Using
their Equation~17, we set

\begin{eqnarray} V(r) = 2\sigma \sqrt{ (\frac{{\cal L}}{{\cal L}_M} - 1)
\ln{\frac{r}{R_0}}},
\end{eqnarray} 
where $\sigma$ is the stellar velocity dispersion of the host galaxy,
which we estimate below; ${\cal L}$ is the quasar
luminosity; $R_0$ is the launch radius of the outflow; and $r$ the
radius at which the velocity of the wind is measured; ${\cal L}_M =
\frac{4\ f_g\ c} {G} \sigma^4$ is a critical luminosity that depends
on the stellar velocity dispersion $\sigma$, the speed of light $c$,
gravitational constant $G$, and the gas fraction $f_g$. For ${\cal
  L}>{\cal L}_M$, radiation pressure may launch a wind. These
equations are appropriate for the limiting case of an optically thick
wind, in which case the interaction is extremely efficient. For these
estimates we assume a launch radius of the wind ($R_0$) of a few 100
pc \citep[the sizes of the circumnuclear molecular disks found in
  low-redshift ULIRGs][ and the lowest value in the AGN feedback
  models of \citealt{ciotti09,ciotti10}]{downes98}, and an outflow
radius, $r$, of 5~kpc, roughly the radius that we spatially resolve.

We have an $L_{bol}$ estimate for J201943$-$364542 from H$\alpha$,
which gives $5 \times 10^{45}$ erg s$^{-1}$. For a fiducial mass of
$5\times 10^{10}$ M$_{\odot}$ in a pressure-supported isothermal
sphere (approximated by the lowest dynamical mass estimate found
 in
 \S\ref{s:additionalLineEmittersTracersOfEnvironmentDynamicalProbes})
, we find a velocity dispersion of $\sigma=\sqrt{M\ G / (5\times R)}$
with stellar velocity dispersion $\sigma$, mass $M$, gravitational
constant $G$, and radius $R$. For higher mass estimates, the
  critical luminosity will also be greater. We assume R=3~kpc, which
gives a velocity dispersion of $\sigma=$210 km s$^{-1}$. The critical
luminosity to launch a wind, ${\cal L}_M$, for this velocity
dispersion and a gas fraction of 10\% is ${\cal L}_M = 3.5\times
10^{46}$ {\rm erg\ s}$^{-1}$.  This suggests that the AGNs in our
sources, unless J201943$-$364542 is atypically weak, do not have
sufficiently powerful quasars to launch fast outflows. We should also
note that the Murray et al. estimate is, strictly speaking, only valid
for buried quasars, whereas the overall low extinction in our sources
(\S\ref{ssec:extinction}) suggests that our galaxies are not very
dusty. Hence, the actual energy transfer from the AGNs to the gas
should be lower than estimated here. For this estimate, we used a very
low value for the fiducial mass. Had we used the average mass of the
HzRGs from the Herge sample instead \citep[][]{seymour07,debreuck10},
$2\times 10^{11}$ M$_{\odot}$, we would have found a circular velocity
of 420 km s$^{-1}$, and a critical luminosity of $5.5\times 10^{47}$
erg s$^{-1}$.

\section{A generic phase in the evolution of massive high-redshift galaxies}
\label{sec:ensemble}

The radio luminosity functions of \citet{Willott2001} and
\citet{Gendre2010} suggest that galaxies with a radio power of
$10^{27-28}$ W Hz$^{-1}$ have co-moving number densities of
$10^{-(7-6)}$~Mpc$^{-3}$ at z$\sim2$, whereas the general population
of massive high-redshift galaxies have densities of a few $10^{-5}$
Mpc$^{-3}$ \citep[e.g.,][for M$_{stellar}\ge 10^{10.5}$
 M$_{\odot}$]{mancini09}. This suggests that phases of radio
activity in this power range could be a generic phase in the evolution
of massive galaxies at these redshifts, if the activity timescales are
short enough, a few $10^7$ yrs. To estimate the duty cycle correction,
we assumed a typical formation epoch of z$\sim$10 for massive high-z
galaxies \citep[e.g.,][]{rocca13}, which implies an age of about
2~Gyrs for massive galaxies at z$\sim2$. The ratio of co-moving number 
densities of radio galaxies and massive galaxies overall would then
imply a duty cycle of about 100, i.e., an activity timescale of a few
$10^7$ yrs, which is consistent with the young ages of high-redshift
radio sources estimated from adiabatic jet expansion models
\citep[][see also \citealt{kaiser97}]{blundell99}, and the empirical
finding of \citet{sajina07} that one-third of the infrared selected
starburst galaxies at z$\sim$2 have extended radio jets with powers of
a few $10^{25}$ W Hz$^{-1}$.

It thus appears  possible that the majority of massive galaxies
experienced such a phase of moderate radio activity if these phases
were sufficiently short \citep[a few $10^7$ yrs, see also][]{venemans07,
nesvadba08}. By studying a sample of radio-selected galaxies,
we might expect to  predominantly probe galaxies in dense
environments \citep[][]{best00,hatch14}. Eight of our sources are
part of the CARLA survey \citep[][]{wylezalek13} with Spitzer, which
measures the density of galaxies with mid-infrared colors consistent
with being at redshifts z$\ge$1.3 around HzRGs. Only one source,
NVSS~J204601$-$335656, has an environment that exceeds the density
of galaxies in the field by more than 3$\sigma$. This suggests that
environment does not  likely   play a dominant role, particularly as 
massive galaxies  generally prefer dense environments 
\citep[e.g.,][]{baldry06}.

\section{Summary}
\label{ssec:summary}

We presented a combined SINFONI and ATCA study of 18 high-redshift
radio galaxies at z$\sim2-3.5$ taken from the MRCR-SUMSS and CENSORS
surveys. Their radio power is in the range of a few $10^{26-27}$ W
Hz$^{-1}$ at 1.4~GHZ, $1-2$ orders of magnitude lower than in the most
powerful high-redshift radio galaxies known, but higher than that 
produced by the most intense high-z starbursts alone. Our goal is to
investigate the ability of moderately powerful radio jets to
accelerate and heat the gas of their host galaxies in order to
regulate the star formation, and the significance of this process in
such a case.

In the near-infrared imaging spectroscopy observations of our sample,
we typically detect faint, unresolved continuum emission around
which we find extended emission-line regions (clearly seen in [OIII]
and H$\alpha$). The kinematic properties of this ionized gas are
diverse among our sample: some sources (e.g., NVSS~J012932$-$385433,
CEN~072) show large-scale and smooth velocity gradients, but of small
amplitude (typically $\lesssim 400$~km s$^{-1}$), while other sources
have very irregular velocity fields (e.g., NVSS~J002431$-$303330). A
common feature of all sources in our sample are their large velocity
dispersions, FWHM~=~$400 - 1000$~km s$^{-1}$. The small ratios of bulk
velocities to velocity dispersions indicate that the observed ionized
gas cannot be in a stable rotating disk, even in cases where smooth
velocity gradients are detected.

Our estimates of ionized gas masses
are in the range of a few $10^{8}$~M$_{\odot}$, which is at least one
order of magnitude less than  was previously found in the most
powerful radio galaxies. For  two sources, we found distinct
emission-line regions in the vicinity of the radio galaxy. Assuming
they are associated with satellite galaxies, we used them to estimate
the dynamical mass of the radio galaxy, and we found a few
$10^{11}$~M$_{\odot}$, which is comparable to the typical mass of
HzRGs \citep[e.g.,][]{Seymour2007, deBreuck2010}. In one source
(NVSS~J201943$-$364542), we find a broad component of H$\alpha$ with
FWHM~$\sim$~8250~km s$^{-1}$. We interpret this as the signature of
the nuclear broad-line region and derive the mass ($\rm M_{BH} \sim
2.1$~M$_{\odot}$) and bolometric luminosity ($5.3 \times 10^{45}$ erg
s$^{-1}$) of the supermassive black hole. This suggests an Eddington
ratio of $\sim$~2~\%.

We explore different possible sources of energy that can explain the
large observed kinetic energy in the ionized gas: transfer from the
radio jet or radiation pressure from the large bolometric luminosity
of the AGNs. We show that an energy transfer from the radio jet to the
ionized gas is a plausible scenario. Our estimates demonstrate that a
fraction of the radio jet power is sufficient to power the kinematics
of the ionized gas. Our observations are in agreement with the
predictions of hydrodynamical models \citep[see, e.g.,][]{Wagner2012}.
  
\section*{Acknowledgments}

We are very grateful to the staff at Paranal for having carried out
the observations on which our analysis is based and to the staff at
the ATCA for their hospitality during our visitor-mode
observations. We thank the anonymous referee for inspiring comments
which helped improve the paper. NPHN wishes to thank G.~Bicknell,
C.~Tadhunter, and J.~Silk for interesting discussions. She also wishes
to thank C.~Harrison for interesting discussions and for pointing out
a missing factor of 1/3 in her previous estimates of warm ionized gas
masses. CC acknowledges support from the Ecole Doctorale Astronomie \&
Astrophysique de l'Ile de France.  Parts of this research were
conducted by the Australian Research Council Centre of Excellence for
All-sky Astrophysics (CAASTRO), through project number CE110001020.
\bibliographystyle{aa}
\bibliography{bibliography,hzrg}

\onecolumn

\begin{figure} \centering
\includegraphics[width=0.8\textwidth]{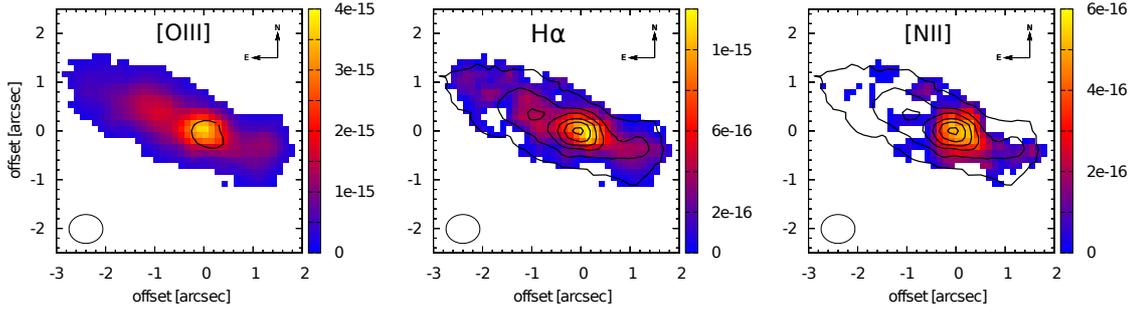}
\caption{{\it Left to right:} Surface brightness maps of
  NVSS~J210626$-$314003 of the [OIII], H$\alpha$, and [NII]. Lines with
  SNR~$\ge$~3.0 are shown.  Contours in the central and right panel
  show the [OIII] morphology to facilitate orientation.\label{fig:J2106}}
\end{figure}

\begin{figure}
\begin{center}
\includegraphics[width=0.8\textwidth]{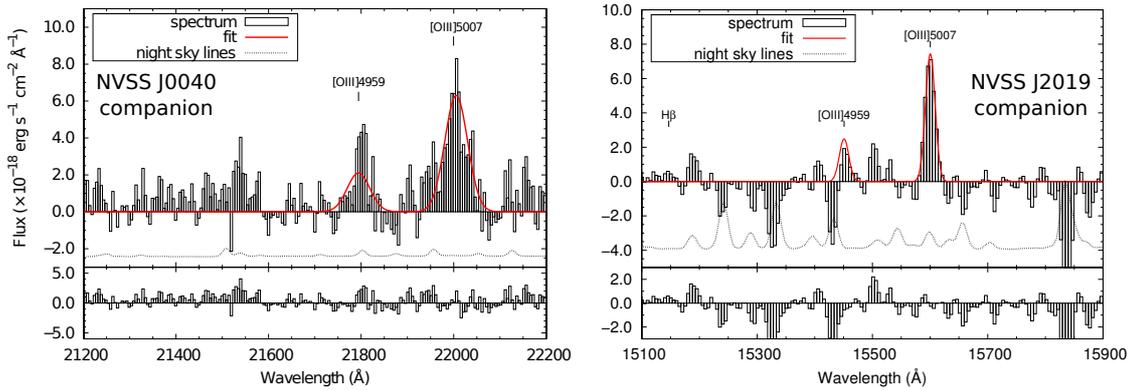}
\caption{{\it Left}: Integrated spectrum of the southern part of the
  emission-line region of NVSS~J004000$-$303333 that appears for
  SNR~$\ge$~3 and that is blueshifted at $z_{\mathrm{south}}^{0040} =
  3.395 \pm 0.001$, as illustrated in Fig.~\ref{fig:J0040_SN3}. {\it
    Right}: Spectrum integrated over a 1.0\arcsec$\times$1.0\arcsec\
  box situated 3.2\arcsec\ to the south-southeast of
  NVSS~J201943$-$364542. The [OIII] emission is detected at 
  redshift $z_{\mathrm{south}}^{2019} = 2.116 \pm
  0.001$.\label{fig:J2019_Spectrum2ndRegion}}
\end{center}
\end{figure}

\begin{figure}
\begin{center}
\includegraphics[width=0.8\textwidth]{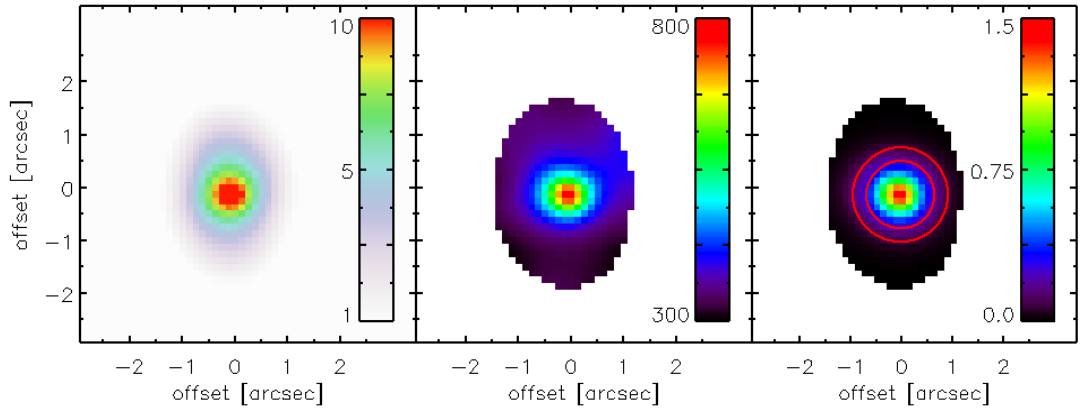}
\caption{Results of our toy models to estimate the impact of an
  unresolved, putative narrow-line region on our estimates of the
  extended gas kinematics. See \S\ref{ssec:toymodels} for
  details. {\it left:} Surface brightness map. {\it center:} FWHMs of
  the cube with extended and narrow-line component. {\it right:}
  Relative increase in line widths between the cube with and without
a  narrow-line region. The red circles show radii corresponding to 1
  and 1.5 times the size of the seeing disk.}
\label{fig:toymodel}
\end{center}
\end{figure}

\begin{figure}
\begin{center}
\includegraphics[width=0.8\textwidth]{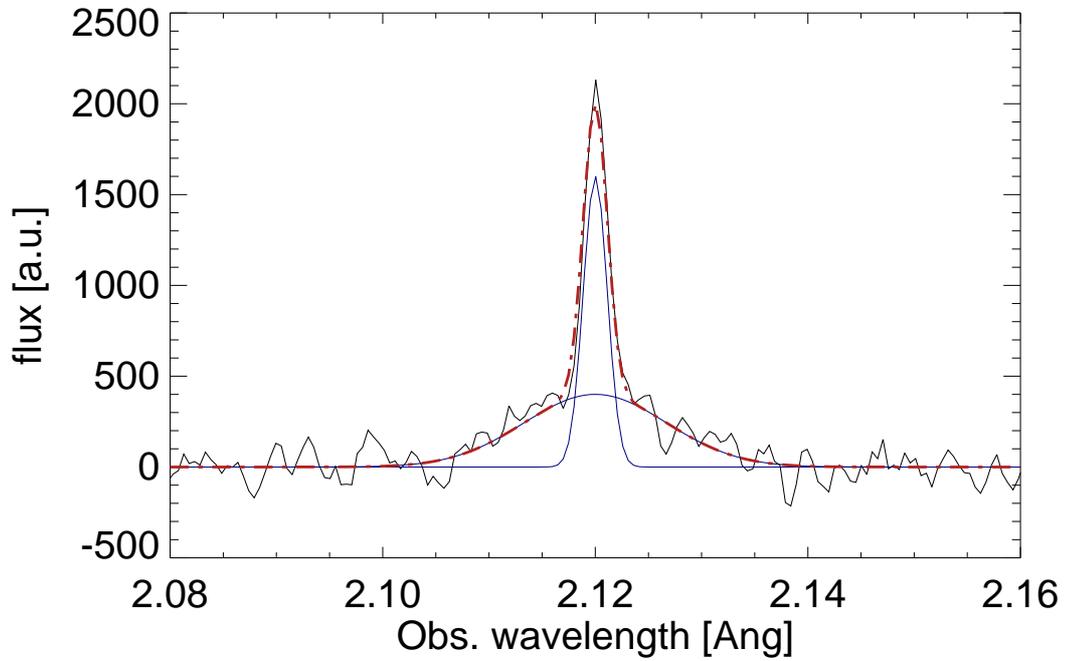}
\caption{Integrated spectrum of the artificial data cube shown in
  Fig.~\ref{fig:toymodel}.  The broad- and narrow-line components are
  clearly distinguishable, and the line profile is more reminiscent of
  bright quasars than the radio galaxies discussed here (see also
  \S\ref{ssec:buitrago}).}
\label{fig:toymodelspec}
\end{center}
\end{figure}

\begin{figure}
\centering
\includegraphics[width=0.8\textwidth]{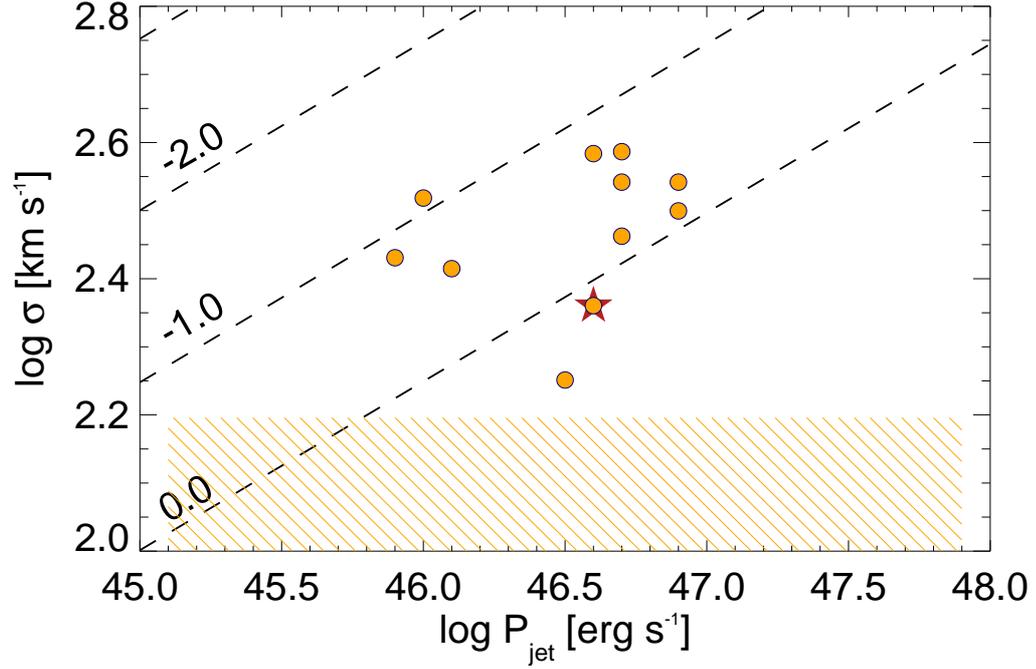}
\label{fig:wagnerfig}
\caption{Gas velocity as a function of kinetic jet power, P$_{jet}$,
  for our galaxies (red dots) and in the model of \citet{Wagner2012} for
  different Eddington ratios (dashed lines). The black star marks
  NVSS~J201943$-$364542, where we have a direct estimate of the Eddington
  ratio. The light orange hatched area marks the velocity
  range in the sample of \citet{buitrago13}, which has no signatures
  of powerful AGNs. This suggests that for jets less powerful than
  ours, kinematic signatures will be very difficult to disentangle
  from other kinematic processes at high redshift.}
\end{figure}

\clearpage

\begin{table}
\caption{Rest-frame UV/optical source properties and observational parameters.\label{tab:obslog}}
\begin{minipage}[t]{\columnwidth}
\centering
\renewcommand{\footnoterule}{}  
\begin{tabular}{lccccccccc}
\hline
Source ID\footnote{Source name in the NVSS catalog. For sources from the CENSORS sample, we give the CENSORS ID in parantheses.} & RA & Dec & $z_{UV}$\footnote{Rest-frame UV spectroscopic redshift of \citet{Bryant2009b} and Johnston et al. (in prep.) for the MRCR-SUMSS sample and of \citet{Brookes2008CENSORS} for the CENSORS sample, respectively.} &  $D_{\rm L}$\footnote{Luminosity distance.} & ToT\footnote{On-source observing time.} & Seeing\footnote{We only list the seeing for galaxies with SINFONI detections.} & K\footnote{K-band magnitude of \citet{Bryant2009a} for the MRCR-SUMSS sample and of \citet{Brookes2008CENSORS} for the CENSORS sample.} \\
                   & (J2000)     & (J2000)     &       & [Gpc] & [min] & [arcsec $\times$ arcsec]         &  [mag]           \\
\hline 
NVSS~J002431$-$303330                 & 00:24:31.80 & -30:33:28.6 & 2.416 &  19.56 & 230 & $0.8 \times 0.7$ & 18.8$\pm$0.1  \\
NVSS~J004000$-$303333                 & 00:40:00.01 & -30:33:32.7 & 3.393 &  29.53 & 180 & $0.7 \times 0.6$ & 19.5$\pm$0.2  \\ 
NVSS~J004136$-$345046                 & 00:41:36.22 & -34:50:46.8 & 2.635 &  21.74 & 180 & \dots & 19.2$\pm$0.3  \\
NVSS~J012932$-$385433                 & 01:29:32.92 & -38:54:34.5 & 2.182 &  17.31 & 225 & $0.9 \times 0.8$ & 18.5$\pm$0.1  \\
NVSS~J030431$-$315308                 & 03:04:31.91 & -31:53:08.4 & 2.250 &  17.89 & 190 & $1.2 \times 1.0$ & 18.7$\pm$0.2  \\
NVSS~J103615$-$321659                 & 10:36:15.26 & -32:16:57.4 & 2.136 &  16.84 & 180 & $0.7 \times 0.6$ & 19.3$\pm$0.2  \\
NVSS~J144932$-$385657                 & 14:49:32.79 & -38:56:57.5 & 2.149 &  16.97 & 220 & $0.9 \times 0.8$ & 19.8$\pm$0.2  \\
NVSS~J201943$-$364542                 & 20:19:43.54 & -36:45:43.2 & 2.128 &  16.69 & 225 & $0.9 \times 0.8$ & 18.4$\pm$0.3  \\
NVSS~J204601$-$335656                 & 20:46:01.08 & -33:56:57.1 & 2.502 &  20.39 & 180 & $0.7 \times 0.6$ & 19.7$\pm$0.4  \\
NVSS~J210626$-$314003                 & 21:06:25.90 & -31:40:01.5 & 2.104 &  16.61 & 180 & $0.7 \times 0.6$ & 18.7$\pm$0.2  \\
NVSS~J233034$-$330009                 & 23:30:34.49 & -33:00:11.5 & 2.675 &  22.14 & 180 & \dots & 17.2$\pm$0.3  \\
NVSS~J234235$-$384526                 & 23:42:35.04 & -38:45:25.0 & 3.507 &  30.74 & 185 & $1.1 \times 1.0$ & 19.0$\pm$0.1  \\
\hline 
NVSS~J094604$-$211508 (CEN~020)       & 09:46:04.55 & -21:15:04.8  & 1.377 & 19.815 & 345  & \dots            & $> 19.4$      \\
NVSS~J094925$-$203724 (CEN~072)\footnote{Coordinates in \citet{Brookes2008CENSORS} are incorrect.}& 09:49:26.00 & -20:37:23.7  & 2.427 & 19.672 & 320  & 1.0$\times$0.9   & 18.88         \\ 
NVSS~J094724$-$210505 (CEN~105)       & 09:47:24.38 & -21:05:02.3  & 3.377 & 29.295 & 440  & \dots            & 20.70         \\
NVSS~J094748$-$204835 (CEN~118)       & 09:47:48.46 & -20:48:35.2  & 2.294 & 18.369 & 295  & \dots            & 19.82         \\
NVSS~J095226$-$200105 (CEN~129)       & 09:52:26.41 & -20:01:07.1  & 2.421 & 19.613 & 305  & 0.9$\times$0.7   & 19.51         \\
NVSS~J094949$-$213432 (CEN~134)       & 09:49:48.77 & -21:34:28.2  & 2.355 & 18.965 & 390  & 1.1$\times$0.9   & 20.24         \\
\hline 
\end{tabular}
\end{minipage}
\end{table}

\begin{table}
\caption{ATCA observing log for the MRCR$-$SUMSS sample.\label{tab:atcalog}}
\begin{minipage}[t]{\columnwidth}
\centering
\renewcommand{\footnoterule}{}  
\begin{tabular}{llllll}
\hline
Source ID     & Date         & TOT \footnote{On-source observing time.}& Secondary Cal.\footnote{Secondary calibrator.} & beam (5.5~GHz)\footnote{Beam size along the major and minor axis, respectively, and position angle measured  from north through east.}  & beam (9.0~GHz)$^c$               \\
              &              & [min]   &                   & [arcsec$\times$arcsec, deg.] & [arcsec$\times$arcsec, deg.]  \\
\hline 
NVSS~J002431$-$303330     &  2012 Jan 28  & 65     & 2357$-$318           & 4.5$\times$1.4, $0$        & 2.9$\times$0.9, $0$        \\
NVSS~J004000$-$303333     &  2012 Jan 28  & 65     & 2357$-$318           & 4.5$\times$1.4, $-1$       & 2.8$\times$0.9, $-1$       \\
NVSS~J012932$-$385433     &  2012 Jan 28  & 65     & 0104$-$408           & 3.6$\times$1.5, $-6$       & 2.3$\times$0.9, $-6$       \\
NVSS~J030431$-$315308     &  2012 Jan 28  & 65     &j0330$-$4014          & 4.2$\times$1.6, $-21$      & 2.7$\times$1.0, $-19$      \\
NVSS~J144932$-$385657     &  2012 Feb 02  & 25     & 1458$-$391           & 6.3$\times$1.5, $-42$      & \dots                      \\
NVSS~J201943$-$364542     &  2012 Feb 02  & 70     & 2054$-$377           & 3.9$\times$1.5, $14$       & 2.6$\times$0.9, $13$       \\
NVSS~J204601$-$335656     &  2012 Feb 02  & 70     & 2054$-$377           & 4.3$\times$1.4, $10$       & 2.8$\times$0.9, $10$       \\
NVSS~J210626$-$314003     &  2012 Feb 02  & 70     & 2054$-$377           & 4.5$\times$1.4, $-7$       & 3.0$\times$0.9, $-8$       \\
NVSS~J234235$-$384526     &  2012 Jan 28  & 65     & 2329$-$384           & 3.7$\times$1.5, $3$        & 2.3$\times$0.9, $4$        \\
\hline 
NVSS~J094604$-$211508 (CEN~020)   & 2012 Jan 28 & 75           & 0925$-$203   & 5.5$\times$1.5, $-0.5$  & 3.4$\times$0.9, $-1$  \\
NVSS~J094925$-$203724 (CEN~072)   & 2012 Jan 28 & 75           & 0925$-$203   & 5.7$\times$1.5, $-0.1$  & 3.5$\times$0.9, $0$  \\
NVSS~J094724$-$210505 (CEN~105)   & 2012 Jan 28 & 75           & 0925$-$203   & 5.5$\times$1.5, $-0.1$  & 3.5$\times$0.9, $0$  \\
NVSS~J094748$-$204835 (CEN~118)   & 2012 Jan 28 & 75           & 0925$-$203   & 6.0$\times$1.5, $-0.7$  & 3.7$\times$0.9, $-1$  \\
NVSS~J095226$-$200105 (CEN~129)   & 2012 Jan 28 & 75           & 0925$-$203   & 6.2$\times$1.5, $-0.7$  & 3.8$\times$0.9, $-1$  \\
NVSS~J094949$-$213432 (CEN~134)   & 2012 Jan 28 & 75           & 0925$-$203   & 5.7$\times$1.5, $-0.5$  & 3.5$\times$0.9, $0$  \\
\hline 
\end{tabular}
\end{minipage}
\end{table}

\begin{table}
\caption{ATCA results. \label{tab:atcaresults}}
\begin{minipage}[t]{\columnwidth}
\centering
\renewcommand{\footnoterule}{}  
\begin{tabular}{lcccccccc}
\hline
Source ID     &  $S_{5.5}$\footnote{Flux measured at 5.5~GHz}  & $S_{9.0}$\footnote{Flux measured at 9.0~GHz} & $\alpha$\footnote{Radio spectral index. (\S\ref{sec:radiocontinuum})}  & $\log$ L$_{500}$\footnote{Radio power at 500~MHz.} & $\log$ L$_{1.4}$\footnote{Radio power at 1.4~GHz in the rest frame.} &  LAS\footnote{Largest angular size, corresponding to the separation between
the two lobes (when detected) or to the deconvolved size for compact sources. For unresolved sources the deconvolved sizes at 5.5~GHz and 9.0~GHz are both given.}      & PA\footnote{Position angle measured from north to east.} \\
              & [mJy]          & [mJy]        &                 & [W Hz$^{-1}$]& [W Hz$^{-1}$]& [arcsec]        & [deg.]    \\
\hline 
NVSS~J002431$-$303330          &  21.1     & 11.1      & $-0.99\pm0.05$  & 27.9 & 27.5 &  16      & 18     \\
NVSS~J004000$-$303333          &  10.0     &  3.4      & $-1.34\pm0.06$  & 28.6 & 27.9 &  15      & 91     \\
NVSS~J012932$-$385433          &  29.2     & 14.3      & $-1.06\pm0.02$  & 28.1 & 27.6 &  $<$1.4  & \dots  \\
NVSS~J030431$-$315308          &  13.4     &  5.9      & $-1.25\pm0.03$  & 28.1 & 27.5 &  $<$3.6  & \dots  \\
NVSS~J144932$-$385657          &  \dots    & \dots     & $-0.99\pm0.05$  & 27.9 & 27.4 &  13.4    & 59     \\
NVSS~J201943$-$364542          &  11.7     & 10.2      & $-1.13\pm0.07$  & 28.0 & 27.5 &  31.4    & 65     \\
NVSS~J204601$-$335656          &  9.8      &  9.1      & $-1.14\pm0.04$  & 28.0 & 27.6 &  3.6     & \dots  \\
NVSS~J210626$-$314003          &  19.8     & 22.4      & $-1.05\pm0.05$  & 28.1 & 27.6 &  50.0    & 98     \\
NVSS~J234235$-$384526          &  6.0      &  2.2      & $-1.38\pm0.03$  & 28.5 & 27.9 &  19.6    & 54     \\
\hline 
NVSS~J094604$-$211508 (CEN~020)&  14.5    & 8.0       & $-1.02\pm0.04$   & 27.3 & 26.8  &   $<$2.9 & 0      \\
NVSS~J094724$-$210505 (CEN~105)&  2.2     & 0.8       & $-1.18\pm0.09$   & 27.5 & 26.8  &  28.4    & 95     \\
NVSS~J094748$-$204835 (CEN~118)&  1.9     & 0.6       & $-1.29\pm0.15$   & 27.6 & 26.9  &   10.    & 0      \\ 
NVSS~J095226$-$200105 (CEN~129)&  2.7     & 1.7       & $-0.82\pm0.01$   & 26.9 & 26.5  &   5.0    & 95     \\
NVSS~J094949$-$213432 (CEN~134)&  2.4     & 1.0       & $-1.08\pm0.09$   & 27.0 & 26.6  &   4.5    & 131    \\
\hline 
\end{tabular}
\end{minipage}
\end{table}

\begin{table}
\caption{Emission-line properties of the MRCR-SUMSS sources.\label{tab:spec1}}
\begin{minipage}[t]{\columnwidth}
\centering
\renewcommand{\footnoterule}{}  
\begin{tabular}{llcccc}
\hline
Source & Line       & $\lambda_0$\footnote{Rest-frame wavelength.} & $\lambda_{\mathrm{obs}}$\footnote{Observed wavelength.} & FWHM\footnote{Full width at half maximum.} & Flux\footnote{Integrated line flux.}          \\ 
& $\quad$    & [\AA]      & [\AA]                 & [km s$^{-1}$]           & [$10^{-16}$erg s$^{-1}$ cm$^{-2}$]  \\ 
\hline 
NVSS~J002431$-$303330 
& H$\beta$        & 4861.3      & 16603.4 $\pm$ 5.3      &  675 $\pm$ 70    &  4.1 $\pm$ 1.0                   \\ 
& $[$OIII$]$      & 4958.9      & 16936.8 $\pm$ 5.4      &  903 $\pm$ 40    & 10.6 $\pm$ 2.1                   \\ 
& $[$OIII$]$      & 5006.9      & 17100.7 $\pm$ 5.4      &  903 $\pm$ 40    & 32.2 $\pm$ 6.4                   \\ 
& $[$NII$]$       & 6548.1      & 22364.6 $\pm$ 7.1      &  220 $\pm$ 25    &  1.6 $\pm$ 0.3                   \\ 
& H$\alpha$       & 6562.8      & 22414.8 $\pm$ 7.1      &  675 $\pm$ 70    & 12.9 $\pm$ 2.6                   \\ 
& H$\alpha$$_{b}$  & 6562.8      & 22414.8 $\pm$ 7.1      & 3250 $\pm$ 300   & 20.8 $\pm$ 4.2                   \\ 
& $[$NII$]$       & 6583.4      & 22485.1 $\pm$ 7.1      &  220 $\pm$ 25    &  4.8 $\pm$ 1.0                   \\ 
\hline 
NVSS~J004000$-$303333 
& $[$OII$]$  & 3727.5               & 16397.3 $\pm$ 4.0  & 900 $\pm$ 90  & 16.5 $\pm$  3.3   \\
& H$\beta$\footnote{Common fit with $[$OIII$]$ with a single redshift and line width.} & 4861.3               & 21385.6 $\pm$ 3.7  & 744 $\pm$ 70  &  5.8 $\pm$  1.2   \\
& $[$OIII$]$                       & 4958.9               & 21814.9 $\pm$ 3.8  & 744 $\pm$ 70  & 22.3 $\pm$  4.5   \\
& $[$OIII$]$                       & 5006.9               & 22026.1 $\pm$ 3.8  & 744 $\pm$ 70  & 67.5 $\pm$ 13.5   \\
\hline 
NVSS~J004000$-$303333~B\footnote{Companion source. See text for details.} 
& $[$OIII$]$$_{\mathrm{south}}$ & 4958.9               & 21794.0 $\pm$ 3.8  & 760 $\pm$ 80  &  1.3 $\pm$ 0.7  \\
& $ [$OIII$]_{\mathrm{south}}$ & 5006.9               & 22005.0 $\pm$ 3.8  & 760 $\pm$ 80  &  3.9 $\pm$ 1.9  \\
\hline 
NVSS~J012932$-$385433 
& H$\beta$       & 4861.3      & 15484.5 $\pm$ 3.5     &  750 $\pm$  80  &  2.0 $\pm$ 0.4  \\
& $[$OIII$]$     & 4958.9      & 15795.4 $\pm$ 3.6     &  909 $\pm$  80  &  8.0 $\pm$ 1.6  \\
& $[$OIII$]$     & 5006.9      & 15948.3 $\pm$ 3.6     &  909 $\pm$  80  & 24.3 $\pm$ 4.9  \\
& $[$NII$]$      & 6548.1      & 20857.4 $\pm$ 4.7     & 1100 $\pm$ 110  &  2.5 $\pm$ 0.5  \\
& H$\alpha$      & 6562.8      & 20904.6 $\pm$ 4.7     &  750 $\pm$  80  &  7.0 $\pm$ 1.4  \\
& H$\alpha$$_b$  & 6562.8      & 20904.6 $\pm$ 4.7     & 3500 $\pm$ 350  & 20.1 $\pm$ 4.0  \\
& $[$NII$]$      & 6583.4      & 20970.2 $\pm$ 4.7     & 1100 $\pm$ 110  &  7.4 $\pm$ 1.5  \\
& $[$SII$]$      & 6716.4      & 21393.9 $\pm$ 4.8     &  800 $\pm$  80  &  2.9 $\pm$ 0.5  \\
& $[$SII$]$      & 6730.8      & 21439.8 $\pm$ 4.8     &  800 $\pm$  80  &  3.1 $\pm$ 0.5  \\
\hline 
NVSS~J030431$-$315308 
& H$\beta$   & 4861.3      & 15773.2 $\pm$ 2.8    & 470 $\pm$ 50 &  2.5 $\pm$ 0.5  \\
& $[$OIII$]$ & 4958.9      & 16089.9 $\pm$ 2.9    & 683 $\pm$ 50 &  9.7 $\pm$ 1.9  \\
& $[$OIII$]$ & 5006.9      & 16245.6 $\pm$ 2.9    & 683 $\pm$ 50 & 29.5 $\pm$ 5.9  \\
& $[$NII$]$  & 6548.1      & 21246.3 $\pm$ 3.8    & 910 $\pm$ 90 &  2.3 $\pm$ 0.5  \\
& H$\alpha$  & 6562.8      & 21294.0 $\pm$ 3.8    & 470 $\pm$ 50 &  5.7 $\pm$ 1.1  \\
& $[$NII$]$  & 6583.4      & 21360.8 $\pm$ 3.8    & 910 $\pm$ 90 &  6.9 $\pm$ 1.4  \\
\hline 
NVSS~J144932$-$385657 
& H$\beta$     & 4861.3      & 15310.4 $\pm$ 3.4    & 350 $\pm$ 35  &  1.1 $\pm$ 0.2 \\
& $[$OIII$]$   & 4958.9      & 15617.8 $\pm$ 3.5    & 420 $\pm$ 40  &  6.4 $\pm$ 1.3 \\
& $[$OIII$]$   & 5006.9      & 15769.0 $\pm$ 3.5    & 420 $\pm$ 40  & 19.5 $\pm$ 3.9 \\
& $[$NII$]$    & 6548.1      & 20622.9 $\pm$ 4.6    & $\le$200      &  0.5 $\pm$ 0.1 \\
& H$\alpha$    & 6562.8      & 20669.2 $\pm$ 4.6    & 350 $\pm$ 35  &  6.6 $\pm$ 1.3 \\
& $[$NII$]$    & 6583.4      & 20734.1 $\pm$ 4.6    & $\le$200      &  1.4 $\pm$ 0.3 \\
\hline 
NVSS~J201943$-$364542 
& H$\beta$                  & 4861.3      & 15165.3 $\pm$ 4.4     &  440 $\pm$ 50  &  $\le$ 0.5     \\
& $[$OIII$]$                & 4958.9      & 15469.8 $\pm$ 4.5     &  540 $\pm$ 30  &  1.5 $\pm$ 0.3 \\
& $[$OIII$]$                & 5006.9      & 15619.5 $\pm$ 4.5     &  540 $\pm$ 30  &  4.5 $\pm$ 0.9 \\
& $[$NII$]$                 & 6548.1      & 20431.6 $\pm$ 5.9     &  150 $\pm$ 20  &  0.4 $\pm$ 0.1 \\
& H$\alpha$                 & 6562.8      & 20477.5 $\pm$ 5.9     &  440 $\pm$ 50  &  2.5 $\pm$ 0.5 \\
& H$\alpha$$_b$             & 6562.8      & 20477.5 $\pm$ 5.9     & 8250 $\pm$ 800 & 35.9 $\pm$ 7.0 \\
& $[$NII$]$                 & 6583.4      & 20541.8 $\pm$ 5.9     &  150 $\pm$ 20  &  1.1 $\pm$ 0.2 \\
\hline 
NVSS~J201943$-$36454~B\footnote{Companion source. See text for details.} 
& $[$OIII$]$$_{\mathrm{south}}$ & 4958.9      & 15450.9 $\pm$ 4.0     & 320 $\pm$ 30   & 0.5 $\pm$ 0.2  \\
& $[$OIII$]$$_{\mathrm{south}}$ & 5006.9      & 15600.4 $\pm$ 4.0     & 320 $\pm$ 30   & 1.6 $\pm$ 0.3  \\
\hline 
NVSS~J204601$-$335656 
& H$\beta$      & 4861.3       & 17010.6 $\pm$ 5.5     &  840 $\pm$ 80 &  1.1 $\pm$ 0.2 \\
& $[$OIII$]$    & 4958.9       & 17352.1 $\pm$ 5.7     &  820 $\pm$ 80 &  3.4 $\pm$ 0.7 \\
& $[$OIII$]$    & 5006.9       & 17520.0 $\pm$ 5.7     &  820 $\pm$ 80 & 10.2 $\pm$ 2.0 \\
& $[$NII$]$     & 6548.1       & 22913.0 $\pm$ 7.5     &  650 $\pm$ 70 &  0.8 $\pm$ 0.2 \\
& H$\alpha$      & 6562.8      & 22964.4 $\pm$ 7.5     &  840 $\pm$ 80 &  5.8 $\pm$ 1.2 \\
& $[$NII$]$     & 6583.4       & 23036.5 $\pm$ 7.5     &  650 $\pm$ 70 &  2.4 $\pm$ 0.5 \\
\hline 
NVSS~J234235$-$384526 
& $[$OII$]$                 & 3727.5       & 16829.6 $\pm$ 5.0    & 1100 $\pm$ 110  &  7.7 $\pm$ 1.5  \\
& H$\beta$                  & 4861.3       & 21947.0 $\pm$ 5.0    &  820 $\pm$  60  &  2.5 $\pm$ 0.5  \\
& $[$OIII$]$                & 4958.9       & 22387.6 $\pm$ 5.2    &  820 $\pm$  60  & 12.5 $\pm$ 2.5  \\
& $[$OIII$]$                & 5006.9       & 22604.3 $\pm$ 5.2    &  820 $\pm$  60  & 37.8 $\pm$ 7.6  \\
\hline 
& $[$OIII$]$$_{\mathrm{blue}}$  & 4958.9       & 22333.5 $\pm$ 5.2    & 1330 $\pm$ 130  &  5.6 $\pm$ 1.1  \\
& $[$OIII$]$$_{\mathrm{blue}}$  & 5006.9       & 22549.7 $\pm$ 5.2    & 1330 $\pm$ 130  & 16.9 $\pm$ 3.4  \\
\hline 
\end{tabular}
\end{minipage}
\end{table}

\begin{table}
\caption{Emission-line properties of the Censors sources. \label{tab:spec2}}
\begin{minipage}[t]{\columnwidth}
\centering
\renewcommand{\footnoterule}{}  
\begin{tabular}{l l c c c c}
\hline
Source & Line       & $\lambda_0$\footnote{Rest-frame wavelength. } & $\lambda_{\mathrm{obs}}$\footnote{Observed wavelength.} & FWHM\footnote{Full width at half maximum, corrected for instrumental resolution.} & Flux\footnote{Integrated line flux.}     \\ 
& $\quad$    & [\AA]      & [\AA]                 & [km s$^{-1}$]           & [$10^{-16}$erg s$^{-1}$ cm$^{-2}$]  \\ 
\hline 
CEN~072 
& H$\beta$       & 4861.3       & 16662.1 $\pm$ 4.0       & 518 $\pm$ 70  &  7.3 $\pm$ 1.6  \\
& $[$OIII$]$     & 4958.9       & 16989.8 $\pm$ 4.0       & 612 $\pm$ 40  &  7.7 $\pm$ 1.5  \\
& $[$OIII$]$     & 5006.9       & 17161.1 $\pm$ 4.0       & 612 $\pm$ 40  & 23.3 $\pm$ 4.5  \\
& $[$NII$]$      & 6548.1       & 22444.4 $\pm$ 5.2       & 518 $\pm$ 70  &  1.9 $\pm$ 0.4  \\
& H$\alpha$      & 6562.8       & 22494.8 $\pm$ 5.2       & 518 $\pm$ 70  & 21.1 $\pm$ 4.2  \\
& $[$NII$]$      & 6583.4       & 22565.4 $\pm$ 5.2       & 518 $\pm$ 70  &  5.8 $\pm$ 1.2  \\
\hline 
CEN~129 
& H$\beta$\footnote{Common fit with H$\alpha$ with a single redshift and line width.}       & 4861.3       & 16637.0 $\pm$ 4.0       & 1107 $\pm$ 30  &  4.4 $\pm$ 1.0  \\
& $[$OIII$]$     & 4958.9       & 16964.1 $\pm$ 4.0       & 635 $\pm$ 30  & 11.0 $\pm$ 2.2  \\
& $[$OIII$]$     & 5006.9       & 17135.2 $\pm$ 4.0       & 635 $\pm$ 30  & 33.4 $\pm$ 6.6  \\
& $[$NII$]$      & 6548.1       & 22408.6 $\pm$ 5.2       & 1107 $\pm$ 50  &  2.8 $\pm$ 0.6  \\
& H$\alpha$       & 6562.8       & 22458.9 $\pm$ 5.2       & 1107 $\pm$ 50  & 12.7 $\pm$ 2.5  \\
& $[$NII$]$      & 6583.4       & 22529.4 $\pm$ 5.2       & 1107 $\pm$ 50  &  8.4 $\pm$ 1.6  \\
\hline 
CEN~134 
& H$\beta$\footnote{Common fit with H$\alpha$ with a single redshift and line width.}        & 4861.3      & 16307.9 $\pm$ 1.9      & 230 $\pm$ 60  &  2.3 $\pm$ 0.4  \\
& $[$OIII$]$      & 4958.9      & 16635.4 $\pm$ 1.9      & 330 $\pm$ 60  & 11.0 $\pm$ 2.2  \\
& $[$OIII$]$      & 5006.9      & 16796.4 $\pm$ 1.9      & 330 $\pm$ 60  & 33.2 $\pm$ 6.6  \\
& $[$NII$]$       & 6548.1      & 21964.4 $\pm$ 2.5      & 230 $\pm$ 50  &  1.2 $\pm$ 0.2  \\
& H$\alpha$       & 6562.8      & 22013.7 $\pm$ 2.5      & 230 $\pm$ 50  &  9.8 $\pm$ 2.0  \\
& $[$NII$]$       & 6583.4      & 22082.8 $\pm$ 2.5      & 230 $\pm$ 50  &  3.7 $\pm$ 0.7  \\
\hline 
\end{tabular}
\end{minipage}
\end{table}

\clearpage
\begin{landscape}
\begin{table*}
\caption{Emission-line imaging: Observational parameters and results.}
\label{tab:lineimaging}
\begin{minipage}[t]{\columnwidth}
\centering
\renewcommand{\footnoterule}{}  
\begin{tabular}{lccccc}
\hline
Source ID         & major axis & minor axis             &   SB [OIII]\footnote{Surface brightness detection limit at $\pm$50\AA\ around the line} & SB H$\alpha^{(a)}$\\
                  &  [arcsec]  &  [arcsec]    &  [$10^{-17}$ erg s$^{-1}$ cm$^{-2}$ arcsec$^{-2}$] & [$10^{-17}$ erg s$^{-1}$ cm$^{-2}$ arcsec$^{-2}$] \\
\hline
NVSS~J002431$-$303330 &    2.5     &      1.5        &  7.5       &  6.4\\
NVSS~J004000$-$303333 &    2.5     &      1.6        &  3.0       &  ...\\
NVSS~J012932$-$385433 &    1.9     &      1.4        &  2.9       &  2.8\\
NVSS~J030431$-$315308 &    1.8     &      1.4        &  5.5       &  2.1\\
NVSS~J144932$-$385657 &    4.4     &      1.5        &  2.3       &  2.4\\
NVSS~J201943$-$364542 &    1.1     &      0.8        &  4.8       &  2.6\\
NVSS~J204601$-$335656 &    1.1     &      0.9        &  5.5       &  5.9\\
NVSS~J210626$-$314003 &    4.9     &      1.7        &  6.8       &  4.1\\
NVSS~J234235$-$384526 &    2.1     &      1.0        &  5.6       &  ...\\
\hline
NVSS~J094925$-$203724 (CEN~072)   & 1.6 & 1.2        &  5.5       &  4.9\\
NVSS~J095226$-$200105 (CEN~129)   & 2.3 & 1.7        &  3.2       &  3.5\\
NVSS~J094949$-$213432 (CEN~134)   & 3.1 & 1.8        &  2.4       &  2.9\\
\hline
\end{tabular}
\end{minipage}
\end{table*}
\end{landscape}

\begin{landscape}
\begin{table*}
\caption{Integrated line properties and related estimates.}
\label{table:ExtinctionElectronDensityMassEnergy}
\begin{minipage}[t]{\columnwidth}
\centering
\renewcommand{\footnoterule}{}  
\begin{tabular}{lcccccccccc}
\hline
ID\footnote{Source ID.} & A(H$\beta$)\footnote{Extinction (in magnitude) at the wavelength of H$\beta$, deduced from the measured
    H$\alpha$/H$\beta$ line ratio and assuming an intrisic Balmer decrement of 2.9.} & ${L}_{H\alpha}$\footnote{H$\alpha$ luminosity, not corrected for extinction.} & $L_{H\alpha}$\footnote{H$\alpha$ luminosity corrected for extinction.} & $L_{[OIII]}$\footnote{[OIII]$\lambda$5007 luminosity} & $n_e$\footnote{Electron density from the [SII]$\lambda$6716/[SII]$\lambda$6731
    line ratio for a temperature $T=10^4$~K. Where the [SII] doublet is too faint or too heavily blended to provide good measurements, we list a fiducidal value of $n_e = 700$ cm$^{-3}$.} & M$_{ion}$\footnote{Mass of warm ionized gas, using the observed line fluxes. } & M$_{ion}^{corr,}$\footnote{Mass of warm ionized gas, using the extinction-corrected line fluxes.} & $\Delta$v\footnote{Velocity offset.} & FWHM\footnote{Full width at half maximum corrected for the instrumental resolution.} & $v/\sigma$   \\
      & [mag]     & [$10^{43}$ erg s$^{-1}$]& [$10^{43}$ erg s$^{-1}$]& [$10^{43}$ erg s$^{-1}$] & [cm$^{-3}$] & [$10^8$ M$_\odot$]& [$10^8$ M$_\odot$] & [km s$^{-1}$] & [km s$^{-1}$] & \\
\hline 
NVSS~J002431$-$303330 & 0.0       &   5.9 &        5.9   & 14.8 & \dots & 3.8   & 3.8        & 350  & 903 & 1.0   \\
NVSS~J004000$-$303333 & ...       & \dots & $\ge$ 18.0\footnote{Lower limit from H$\beta$ for A$_{\rm V}=0$} & 70.5 & \dots & \dots & $\ge$ 11.6  & 350  & 744 & 0.6    \\
NVSS~J012932$-$385433 & 0.7       &   2.5 &        4.4    &  8.7 &  750  & 1.1   & 2.0        &  350   & 909  & 0.5  \\
NVSS~J030431$-$315308 & 0.0       &   2.2 &        2.2    & 11.3 & \dots & 1.4   & 1.4        &  250   & 683  & 0.5  \\
NVSS~J144932$-$385657 & 1.9       &   2.3 &       10.9    &  6.7 & \dots & 1.4   & 7.1        &  800   & 420  & 2.3  \\
NVSS~J201943$-$364542 & $\ge$ 1.7 &   0.8 & $\ge$  3.6    &  1.5 & \dots & 0.6   & $\ge$ 2.4  &  300   & 540  & 0.7  \\
NVSS~J204601$-$335656 & 0.0       &   2.9 &       2.9     &  5.1 & \dots & 1.8   & 1.8        &  180   & 820  & 0.3  \\
NVSS~J210626-314003   & 0.0       &   4.9 &        4.9    & 20.8 &  500  & 3.2   & 8.3        &  600   & 370  & 1.9  \\
NVSS~J234235$-$384526 & ...       & \dots & $\ge$ 12.7$^k$& 42.7 & \dots & \dots & $\ge$ 5.9  &  450   & 820  & 0.7  \\
\hline 
NVSS~J094925$-$203724 (CEN~072)    & 0.0  & 9.8   & 9.8 & 10.8 & \dots & 6.3   & 4.5         &  300   & 612  & 0.6  \\
NVSS~J095226$-$200105 (CEN~129)    & 0.0  & 5.8   & 5.8 & 15.4 & \dots & 3.8   & 2.7         &  400   & 635  & 0.8  \\
NVSS~J094949$-$213432 (CEN~134)    & 1.1  & 4.2   & 6.1 & 14.3 & \dots & 2.8   & 2.8         &  170   & 330  & 0.6  \\
\hline 
\end{tabular}
\end{minipage}
\end{table*}
\end{landscape}

\begin{landscape}
\begin{table*}
\caption{Kinetic energy of the jet and gas.}
\label{table:Energy}
\begin{minipage}[t]{\columnwidth}
\centering
\renewcommand{\footnoterule}{}  
\begin{tabular}{lccccccc}
\hline
ID      & E$_{bulk}$\footnote{Bulk kinetic energy of the warm ionized gas.} & E$_{turb}$\footnote{Turbulent energy of the warm ionized gas.} & $E_{bulk}/(E_{turb}+E_{bulk})$ & $log L_{jet}^{Willott99,}$\footnote{Mechanical jet power using the calibration of \citet{Willott1999}.}  & $log L_{jet}^{Cavagnolo10,}$\footnote{Kinetic power of the radio jet estimated with the calibation of \citet{Cavagnolo2010}.} & $\tau$\footnote{Fiducial jet age for an advance speed of 0.1$c$.} & $E_{jet}$ \\
        & [$10^{56}$ erg]& [$10^{56}$ erg] &               & [erg s$^{-1}$] & [erg s$^{-1}$] & [$10^5$ yrs] & [$10^{59}$ erg]  \\
\hline
NVSS~J002431$-$303330   &  2.4           & 7.3        & 0.25 & 46.6  & 46.6 & 10.2 & 13    \\ 
NVSS~J004000$-$303333   & $\ge$ 3.5      & $\ge$ 30.8 & 0.1  & 47.0  & 46.9 & 9.0  & 28    \\ 
NVSS~J012932$-$385433   &       0.8      &     13.3   & 0.06 & 46.6  & 46.7 & 1.0  & 1.3   \\ 
NVSS~J030431$-$315308   &       0.4      &      3.5   & 0.1  & 46.7  & 46.7 & 2.4  & 3.8   \\ 
NVSS~J144932$-$385657   &      11.3      &     7.4    & 0.6  & 47.1  & 46.5 & 9.1  & 36    \\ 
NVSS~J201943$-$364542   & $\ge$ 0.13     &$\ge$ 3.9   & 0.03 & 46.7  & 46.6 & 21.3 & 34    \\ 
NVSS~J204601$-$335656   &       0.14     &      8.8   & 0.02 & 46.6  & 46.7 & 2.1  & 2.6   \\ 
NVSS~J234235$-$384526   & $\ge$ 4.2      &$\ge$ 18.6  & 0.2  & 47.0  & 46.9 & 11.7 & 37    \\ 
\hline 
NVSS~J094925$-$203724 (CEN 072)  & 0.2  & 8.1   & 0.02 & 45.7  & 46.1 & 4.9  & 0.8   \\ 
NVSS~J095226$-$200105 (CEN~129)  & 1.0  & 4.8   & 0.2  & 45.5  & 45.9 & 2.9  & 0.3   \\ 
NVSS~J094949$-$213432 (CEN~134)  & 0.2  & 2.6   & 0.09 & 45.7  & 46.0 & 3.5  & 0.6   \\ 
\hline 
\end{tabular}
\end{minipage}
\end{table*}
\end{landscape}

\end{document}